\documentclass{aa}  

\usepackage{amsmath}
\usepackage{graphicx}
\usepackage{changepage}
\usepackage{listings}
\usepackage{siunitx}
\usepackage{natbib}
\usepackage{hyperref}
\usepackage{multirow}
\usepackage{float}
\usepackage{placeins}
\usepackage{pdflscape}  
\usepackage{longtable}  
\usepackage{lastpage}
\usepackage{txfonts}
\usepackage{xcolor}

\begin{document}

   \title{Hydrodynamical mass-loss rates for very massive stars}

   \subtitle{I. Investigating the wind kink}

\author{{Gautham N. Sabhahit\inst{\ref{AOP}}}
    \and 
       {Jorick S. Vink\inst{\ref{AOP}}}
    \and
      {Andreas\,A.\,C. Sander\inst{\ref{ARI}}}
    }

\institute{
   {Armagh Observatory and Planetarium, College Hill, Armagh BT61 9DG, N. Ireland\label{AOP}}
   \and
   {Zentrum f{\"u}r Astronomie der Universit{\"a}t Heidelberg, Astronomisches Rechen-Institut, M{\"o}nchhofstr. 12-14, 69120 Heidelberg, Germany\\\label{ARI}}
   \email{gauthamns96@gmail.com}
             }


 
  \abstract
   {Radiation-driven winds are ubiquitous in massive stars, but in very massive stars (VMSs), mass loss dominates their evolution, chemical yields, and ultimate fate. Theoretical predictions have often relied on extrapolations of O-star prescriptions, likely underestimating true VMS mass-loss rates. In the first of a series of papers on VMS wind properties, we investigate a feature predicted by Monte Carlo (MC) simulations: a mass-loss `kink' or upturn where the single-scattering limit is breached and winds transition from optically thin to optically thick. We calculated hydrodynamically consistent  wind-atmosphere models in non-local thermodynamic equilibrium using the $\texttt{PoWR}^\textsc{hd}$ code, with a grid spanning $40-135\,M_\odot$ and $12-50\,\mathrm{kK}$ at fixed $\log(L_\star/L_\odot) = 6.0$ and solar-like metallicity with $Z=0.02$. Our models confirm the existence of the kink, where the wind optical depth crosses unity and spectral morphology shifts from O-star to WNh types. The predicted location of the kink coincides with the transition stars in the Galactic Arches Cluster and reproduces the model-independent transition mass-loss rate of $\log(\dot{M}_\mathrm{trans}) \approx -5.16$ from Vink \& Gräfener (2012). For the first time, we locate the kink at $\Gamma_\mathrm{e} \approx 0.43$ ($M_\star \approx 60\,M_\odot$) without relying on uncertain stellar masses. Above the kink, mass-loss rates scale much more steeply with decreasing mass (slope $\sim$10), in qualitative agreement with MC predictions. We additionally identified two bistability jumps in the mass loss driven by Fe ionisation shifts: The first is from \ion{Fe}{iv}$\,\rightarrow\,$\ion{Fe}{iii} near $25\,\mathrm{kK}$, and the second is from \ion{Fe}{iii}$\,\rightarrow\,$\ion{Fe}{ii} near $15\,\mathrm{kK}$. Our models thus provide the first comprehensive confirmation of the VMS mass-loss kink while establishing a mass-loss relation with complex mass and temperature dependencies with consequences for stellar evolution, chemical yields, and the black hole mass spectrum.
   }

   \keywords{Stars: atmospheres -- stars: massive -- stars: mass-loss -- stars: winds, outflows -- stars: Wolf-Rayet 
               }

   \maketitle
%

\section{Introduction}
\label{sec: Introduction}

Winds from massive O-type stars in the range of $20-60\,M_{\odot}$ are a key ingredient in stellar evolutionary models, shaping both their structure and their final fate. However, for the most massive stars in the Universe with initial masses above $100\,M_\odot$ -- the so-called very massive stars (VMSs) -- mass loss is not merely important: It is the dominant factor that determines their evolution and ultimate fate \citep{Hirschi2015, Kohler2015}.

O-star winds have long been described by the optically thin line-driving theory of \citet{LS1970} and \citet[][hereafter CAK]{CAK1975}. Evolutionary models for VMSs have traditionally extrapolated these prescriptions into higher masses. However, such extrapolations likely underestimate the true VMS mass-loss rates. \citet{Vink2011} using Monte Carlo (MC) simulations predicted an upturn or a `kink' in the mass-loss–luminosity relation, associated with the Eddington parameter $\Gamma \propto L/M$. This kink suggests that winds strengthen dramatically and become optically thick; that is, the wind optical depth, $\tau_{F}$, crosses unity, as stars approach the Eddington limit.

Concurrently with these VMS wind predictions, the widely assumed stellar upper-mass limit of $100-150\,M_{\odot}$ \citep{Weidner_Kroupa2004, Figer2005,  Oey_Clarke2005} was revised to $\sim$200-300\,$M_{\odot}$ \citep{Crowther2010, Bestenlehner2011, Mart15, Brands22, Kalari22}. These values, however, are based on present-day masses extrapolated to the zero-age main sequence (ZAMS) using O-star rates, and may thus be underestimated. If VMS mass loss is indeed stronger than CAK extrapolations suggest, the intrinsic upper-mass limit may extend significantly higher -- perhaps up to $1000\,M_{\odot}$ \citep{Vink2018, Sabhahit2022, Higgins2022}. A closely related issue concerns the formation of VMSs \citep{Krumholz2015} and the upper-mass limit, which may be regulated by the strength of stellar winds in environments of different metallicity.

In the local Universe, the hydrogen-rich class of Wolf-Rayet stars of the nitrogen (N) sequence (WNh stars) in young ($\sim 1-2$ Myr) massive clusters, which exhibit strong \ion{He}{II} emission, are our best candidates for VMSs \citep{Massey1998, Dekoter1998, Martins2008, Crowther2016, Wofford2014, Upadhyaya2024}.
Such VMSs are not only key to stellar evolution theory but also to astrophysical environments, since they may drive the chemical enrichment of globular clusters \citep{Vink2018, Higgins2023, Higgins2025, Gieles2025} and also provide crucial feedback at high redshift \citep{Senchyna2024, Vink23, Charbonnel2023, Cameron2023, Isobe2023, Topping2024, Curti2025, Ji2025}. 
Predicting their wind yields and final masses requires reliable mass-loss prescriptions, especially across the transition from optically thin ($\tau_F < 1$) O-star winds to optically thick ($\tau_F > 1$) WNh winds \citep{Vink2012}.

The \citet{Vink2011} kink approximately coincides with the wind efficiency parameter $\eta = \dot{M} v_\infty c / L$ approaching unity. Above this threshold, the predicted mass-loss scaling with $\Gamma$ steepens from approximately two to about five, far stronger than a CAK-style extrapolation. While CMFGEN analyses from the VLT-FLAMES Tarantula Survey \citep{Best2014} provide empirical support, several studies still neglect the $\eta \approx \tau_F \approx 1$ criterion emphasised by \citet{Vink2012}.

Given the upturn or kink predicted by the MC simulations, wind calculations using a different method and code are important to independently verify the MC results, which motivates new predictions of VMS mass-loss rates both below and above the potential upturn or kink. Here we present new computations with the hydrodynamically consistent branch of the Postdam Wolf-Rayet (PoWR) code \citep{Sander2017}, as recently applied to the study of late-type WNh stars \citep{Lefever2025} and some of the most massive stars in the R136 cluster in the Large Magellanic Cloud (LMC) \citep{Sabhahit_VMS2025}. These kinds of hydro-models are also ideal tools to study the transition from absorption-line dominated O-star spectra to emission-line dominated WNh spectra as the objects approach the Eddington limit. Along this transition, in particular, the equivalent width of the Wolf-Rayet (WR) He {\sc ii} emission feature increases \citep[see][]{Vink2011, CW2011}.

The paper is organised as follows: In Sec.~\ref{sec: Methods}, we detail the basics of our hydrodynamical wind-atmosphere models and the necessary inputs and outputs. In Secs.~\ref{sec: rad_acceleration} and \ref{sec: results}, we present the radiative acceleration and mass-loss properties of VMSs, discussing the mass-loss kink feature and bistable behaviour at cooler temperatures. We further test our mass-loss predictions against the transition mass-loss rate in the Arches Cluster. A brief comparison to previous VMS wind investigations is presented in Sec.~\ref{sec: discussion}, and we conclude in Sec.~\ref{sec: conclusions}.

\section{$\texttt{PoWR}^\textsc{hd}$ atmosphere models}
\label{sec: Methods}

We utilised the hydrodynamical branch of the non-local thermodynamic equilibrium (non-LTE) stellar atmospheric code $\texttt{PoWR}$ \citep{Grafener2002, HG2003, GH2005, GH2008, Sander2015diss, Sander2015, Sander2017} to build our model atmosphere grid. $\texttt{PoWR}$ models a one-dimensional, spherically symmetric, expanding atmosphere by solving the equations of radiative transfer, statistical and thermal equilibrium, and continuity. The atmosphere computational framework has long relied on the Accelerated Lambda Operator iteration \citep[for details, see][]{Hamann85, Hamann86} to solve these fully coupled non-linear equations across both spatial and frequency domains. Radiative transfer calculations are performed using the co-moving frame approach, in which the opacity and emissivity profile functions become isotropic \citep{Mihalas1975}. The code takes relevant stellar and wind parameters as inputs and computes the emergent synthetic spectra. The input stellar parameters include luminosity, temperature (or radius), log$g$ (or mass), and chemical abundances (for a detailed input parameters, see $\mathrm{Sec.}\,\ref{sec: inputs_Powr_HD}$). The wind parameters are specified by the mass-loss rate, $\dot{M}$, and the velocity stratification, $\varv(r)$, which is typically prescribed using a standard $\beta$-type velocity law of the form $\varv(r) \sim (1 - R_\star/r)^\beta$.

The fundamental advance and novelty of the $\texttt{PoWR}^\textsc{hd}$ models over the base code stems from solving the stationary hydrodynamic equation of motion (cf. Eq.~\eqref{eq: hydro_eq}).  The code balances the time-independent advective acceleration term ($\varv \mathrm{d}\varv/\mathrm{d}r$), against the various forces in the wind-atmosphere, such as gravity and the gas and radiation pressure gradients at each atmospheric layer, thereby integrating hydrodynamics into the overall iterative procedure of the atmosphere code. By solving the wind hydrodynamics, the models gain predictive power, as the mass-loss rate and the wind velocity stratification are no longer prescribed inputs. These quantities instead emerge as predicted outputs, calculated self-consistently from the atmospheric force balance at each iteration step.

Such next-generation hydrodynamically consistent atmosphere models therefore achieve what earlier codes could not: They simultaneously derive consistent wind structure and mass loss, and generate synthetic spectra through unified non-LTE atmosphere-hydrodynamics calculations, enabling empirical determination of fundamental stellar and wind parameters \citep[see e.g.][]{Sander2017, Sabhahit_VMS2025}. The numerical implementation details and general input specifications are documented in detail in \citet{Sander2015diss, Sander2017, Sander2020a}. Below, we briefly summarise the required inputs to build our model grid and the relevant outputs.

\subsection{Key input parameters}
\label{sec: inputs_Powr_HD}

In this work, we isolated the effects of stellar mass and inner boundary temperature on VMS wind properties. While a comprehensive parameter space exploration and theoretical mass-loss predictions for the most-massive WNh stars based on $\texttt{PoWR}^\textsc{hd}$ models will follow in a forthcoming work, the present study maintains all parameters except mass and temperature at the fixed values detailed below.

 We fixed the luminosity to a constant value of $\log(L_\star/L_\odot) = 6.0$ throughout our grid. The inner boundary radius, $R_\star$, of our models is defined at a specific Rosseland continuum optical depth, $\tau_\mathrm{R,cont}$, of $20$. The effective temperature, $T_\star$, at the inner boundary is varied between $12-50\,\mathrm{kK}$, primarily encompassing the temperature range characteristic of observed WNh stars and luminous blue variables (LBVs) at this luminosity. The stellar mass range extended from approximately $40-135\,M_\odot$, with the precise range depending on $T_\star$.\footnote{Table~\ref{tab: wind_properties} in Appendix~\ref{appendix: summary} provides a tabulation of the explored parameter space along with relevant wind predictions.} The outer boundary was set to $R_\mathrm{out} = 1000\,R_\star$.

For the chemical abundances, we adopted ZAMS values following solar-scaled patterns from \citet{GS98}, with a total metal mass fraction of $Z = 0.02$ (Appendix~\ref{appendix: summary} details the mass fraction breakdown for individual elements, along with levels and lines per ion). The hydrogen (H) mass fraction was fixed at $X=0.7$, while helium (He) followed from the unity constraint $X+Y+Z=1$.

Density inhomogeneities -- ubiquitous in massive star winds -- were incorporated through the micro-clumping formalism of \citet{Hillier2003}, with the clumping onset occurring at a specified Rosseland optical depth \citep[see][for the implementation]{Sander2017}. The clumping factor, $D_\mathrm{cl}$, was increased from unity (smooth wind) to an asymptotic value, $D_\mathrm{cl,\infty} = 10$, with the onset at Rosseland optical depth $\tau_\mathrm{R} = 0.1$. This choice of onset is  motivated by multi-wavelength spectral fits to archetypal VMSs in the Tarantula Nebula, specifically R136a1 (WN5h) and the binary system R144 (WN5/6h+WN6/7h) using $\texttt{PoWR}^\textsc{hd}$ models \citep{Sabhahit_VMS2025}. Turbulent pressure contributions were included through adoption of a radially constant turbulent velocity $\varv_\mathrm{turb} = 70\,\mathrm{km\,s^{-1}}$. Such high values are motivated from time-dependent, 2D simulations of massive O-star atmospheres, where substantial photospheric turbulence of the order $30-100\,\mathrm{km\,s^{-1}}$  is expected to originate from the hot iron (Fe) bump \citep{Debnath2024}. 

\subsection{Output quantities}
\label{sec: outputs_Powr_HD}

The $\texttt{PoWR}^\textsc{hd}$ models enforce the stationary hydrodynamic equation of motion throughout the atmosphere:
\begin{equation}
\begin{split}
\varv(r)\dfrac{\mathrm{d}\varv(r)}{\mathrm{d}r} = - \dfrac{1}{\rho(r)}\dfrac{\mathrm{d}P(r)}{\mathrm{d}r} - \dfrac{GM_\star}{r^2} + a_\mathrm{rad}(r),
\end{split}
\label{eq: hydro_eq}
\end{equation}
where $P(r)$ is the combined gas and turbulent pressure, while $\rho(r)$ and $\varv(r)$ denote the mass density and velocity stratifications which are linked to the mass-loss rate $\dot{M}$ through the continuity equation: $\dot{M} = 4\pi r^2 \rho(r) \varv(r)$. The total radiative acceleration $a_\mathrm{rad}(r)$, arising from momentum transfer on transition lines, scattering, and other continuum processes, fundamentally couples the hydrodynamic equation to the radiation field and population numbers.

Equation~\eqref{eq: hydro_eq} fundamentally represents the force balance equation, with the advective acceleration term on the left-hand side counterbalanced by the cumulative atmospheric forces on the right -- gravity and the gradients of gas, radiation, and turbulent pressure. The apparent simplicity of Eq.~\eqref{eq: hydro_eq} is however misleading. The complexity lies masked within the radiative acceleration term $a_\mathrm{rad}(r)$, given by
\begin{equation}
\begin{split}
a_\mathrm{rad}(r)  = \dfrac{1}{c} \int_0^\infty \varkappa_\nu(r) F_\nu(r) \, \mathrm{d}\nu = \dfrac{\varkappa_\mathrm{F}(r) F(r)}{c},
\end{split}
\label{eq: rad_accl}
\end{equation}
where $\varkappa_\mathrm{F}(r)$ and $F(r)$ represent the flux-weighted mean opacity and total flux at radius $r$, respectively. The quantity $\varkappa_\mathrm{F}(r)$ cannot be determined locally; it must instead incorporate the detailed frequency-dependent opacity, which depends on radiation and velocity fields, along with ionisation and excitation states of absorbing ions throughout the atmosphere. Such complexity demands a brute-force, iterative global integration to achieve self-consistent hydrodynamic atmosphere models.

The flux-weighted mean opacity calculation is however essential because it accounts for Doppler effects in the expanding wind and ensures accurate radiative acceleration through conservation of the flux integral in Eq.~\eqref{eq: rad_accl}. Unlike the classical Thomson acceleration -- well constrained at model initialisation, particularly when stellar temperatures are high enough for complete H and He ionisation -- the total radiative acceleration is calculated iteratively. The radial stratification of this total radiative acceleration from converged models is examined in Sect.~\ref{sec: rad_acceleration}.

Using the continuity equation in Eq. \eqref{eq: hydro_eq} and rearranging the hydrodynamic equation  such that all terms containing the velocity gradient are grouped together on one side, we can show that the hydrodynamic equation features a critical point at radius $R_\mathrm{crit}$\citep[see derivation in][]{Sander2017}. The critical point in our model occurs where the flow velocity equals the isothermal sound speed corrected for turbulent velocity\footnote{This is because the radiative acceleration $a_\mathrm{rad}$ is re-calculated every iteration as a function of radius. This is unlike CAK-type approaches where the $a_\mathrm{rad}$ is parametrised as a function of radius and velocity gradient, and the resulting CAK-critical point does not coincide with the sonic point.}, so that the critical radius $R_\mathrm{crit}$ roughly coincides with the sonic point radius $R_\mathrm{sonic}$. Each hydrodynamic update recalculates the velocity stratification throughout the atmosphere via inward and outward integration from $R_\mathrm{crit}$. We therefore predict the complete velocity stratification $\varv(r)$, including the terminal velocity at the outer boundary: $\varv_\infty = \varv(r = R_\mathrm{out})$. Mass-loss rate updates preserve the inner boundary condition to maintain the specified $\tau_\mathrm{R,cont}$.

From the predicted $\dot{M}$ and $\varv_\infty$, the wind efficiency parameter is given by:
\begin{equation}
\begin{array}{c@{\qquad}c}
\eta = \dfrac{\dot{M}\varv_\infty}{L_\star/c}.
\end{array}
\label{eq: eta_wind_efficiency}
\end{equation}
The parameter $\eta$ quantifies the ratio of wind momentum to available photon momentum. For powerful winds characteristic of WNh and classical WR stars, where multiple photon scattering becomes significant, $\eta$ can easily exceed the single-scattering limit of unity.

Another important wind-related quantity outputted by our models is the wind optical depth, specifically, the flux-weighted mean wind optical depth at the sonic point. It is calculated by integrating the product of flux-weighted mean opacity and density inward from the outer boundary:
\begin{equation}
\begin{array}{c@{\qquad}c}
\tau_{F}(r_\mathrm{sonic}) =\displaystyle\int^{\infty}_{r_\mathrm{sonic}} \varkappa_F(r) \rho(r) \mathrm{d}r.
\end{array}
\label{eq: wind_optical_depth}
\end{equation}
Since we obtain a detailed solution to the radiative transfer problem, we predict the flux-weighted mean opacity and therefore output the wind optical depth.

Finally, converged models generate emergent synthetic spectra through formal integration in the observer's frame, incorporating relevant broadening mechanisms. We additionally computed ionising photon fluxes (photons per second) by integrating the emergent spectrum shortward of relevant continuum edges, for instance, the Lyman continuum edge yields the H-ionising flux. These predicted quantities from our models are detailed below, while Appendix~\ref{appendix: summary} provides a comprehensive table of wind-related outputs from our $\texttt{PoWR}^\textsc{hd}$ models.

\section{Radiative acceleration}
\label{sec: rad_acceleration}

\begin{figure}
    \includegraphics[width = \columnwidth]{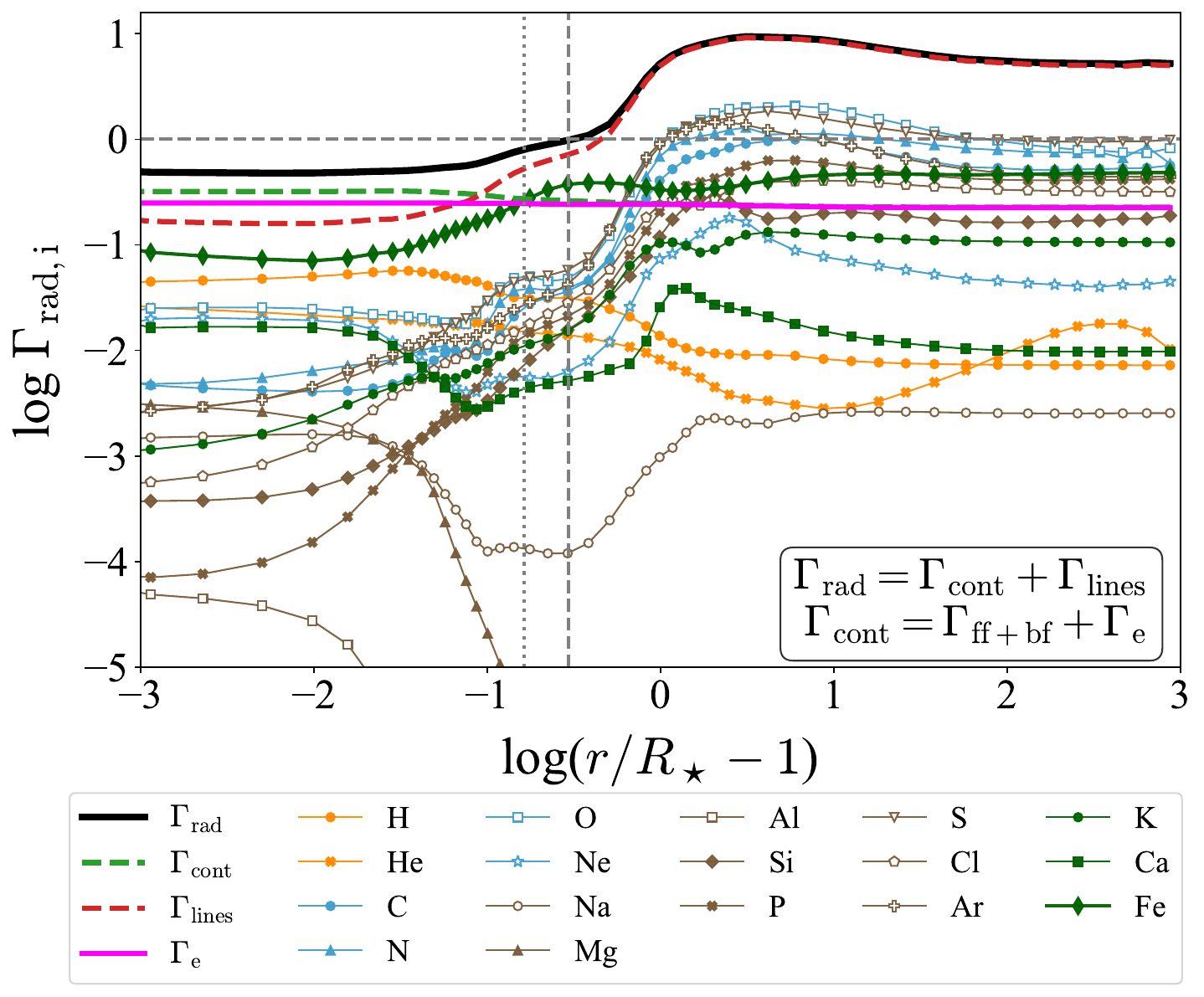}
    \caption{Total radiative acceleration and its contributions from continuum processes and line transitions. Decomposition of the total radiative acceleration (solid black) into continuum (dashed green) and line (dashed red) components along with individual contributions from constituent elements in the atmosphere (coloured lines with symbols, see legend). Each of these element contributions consists of the sum of their line and continuum accelerations. All accelerations are normalised to gravity. The dotted and dashed vertical grey lines mark the sonic and critical points, $R_\mathrm{sonic}$ and $R_\mathrm{crit}$, respectively. The model stellar parameters are $\log(L_\star/L_\odot) = 6.0$, $M_\star = 105\,M_\odot$, $T_\star = 35\,\mathrm{kK}$, $X = 0.7$, and $Z = 0.02$.}
    \label{fig: rad_accl}
\end{figure}

\begin{figure}
    \includegraphics[width = \columnwidth]{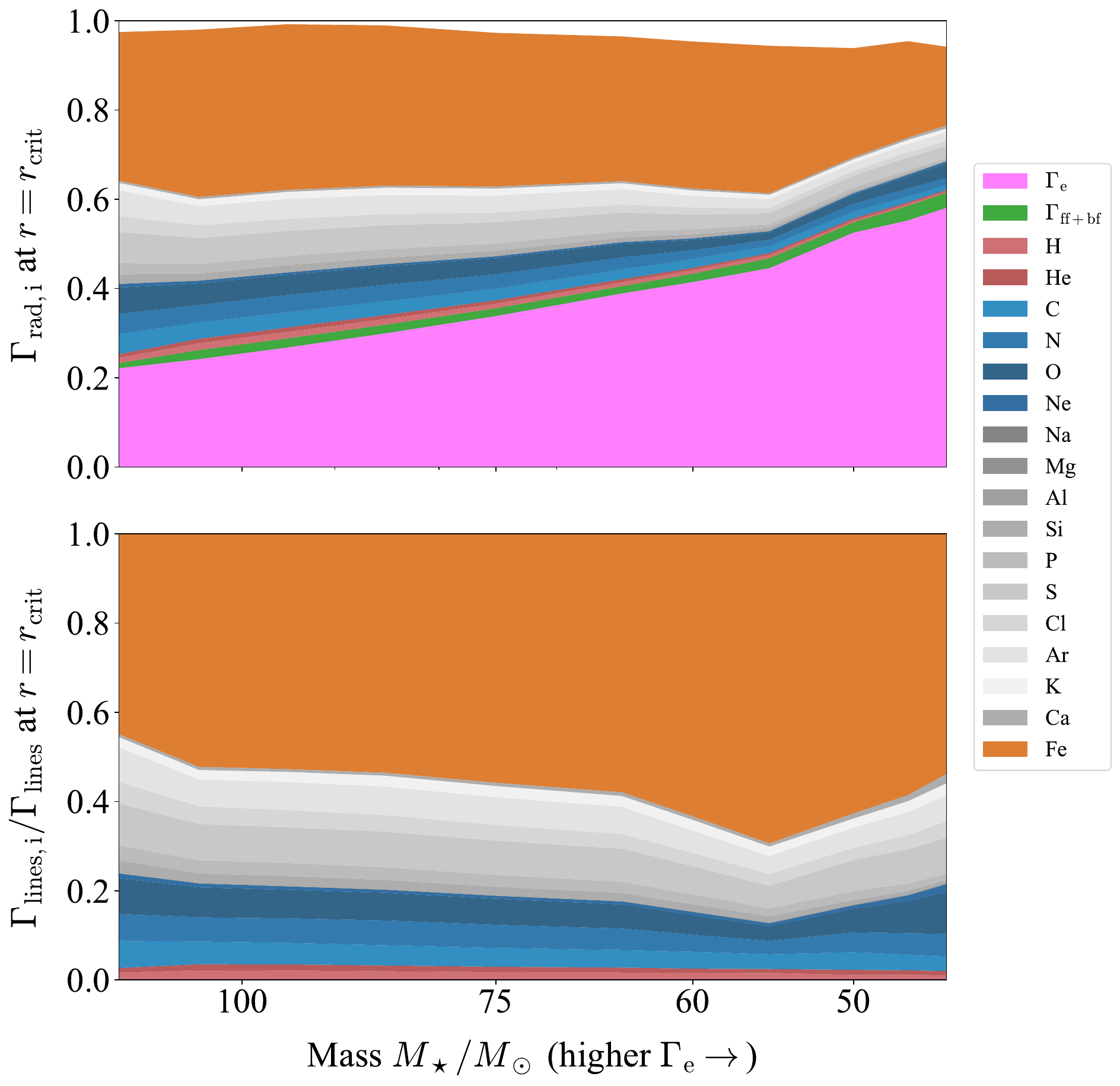}
    \caption{Contributions from electron scattering, free-free and bound-free continuum transitions, and line transitions of individual elements at the critical point, shown as a function of mass. The contributions are normalised to gravity and expressed as Eddington parameters. \textit{Top:} Absolute contributions from different opacity sources. \textit{Bottom:} Relative line contributions from different elements. The model sequence corresponds to a temperature of $T_\star = 35\,\mathrm{kK}$.}
    \label{fig: relative_contribution_total}
\end{figure}

A major advance in our understanding of hot-star winds over recent decades has been recognising the importance of proximity to the Eddington limit \citep{GH2008, Vink2011, Sander2020b}. Yet predicting a priori how metallicity and other stellar parameters affect VMS winds remains challenging, which is precisely why detailed radiative transfer calculations, such as those performed here are essential.

Nevertheless, we can infer qualitative trends to first order from the force balance in stellar winds \citep[see][for a didactic study]{Sabhahit_VMS2025}. The balance of relevant forces in massive star winds is characterised by the total radiative Eddington parameter, defined as the ratio of total radiative acceleration to gravity:
\begin{equation}
\begin{split}
\Gamma_\mathrm{rad}(r)   = \dfrac{a_\mathrm{rad}(r)}{GM_\star/r^2} =  \dfrac{\varkappa_{F}(r)L(r)}{4\pi GcM_\star} \approx \dfrac{\varkappa_{F}(r)L_\star}{4\pi GcM_\star},
\end{split}
\label{eq: eddington_parameter}
\end{equation}
where the radial variation in total $\Gamma_\mathrm{rad}$ primarily reflects changes in the flux-weighted mean opacity. The luminosity $L(r)$ technically varies with radius $r$, since a fraction of the photon energy is used to lift the material out of the stellar gravitational potential and to accelerate it to a certain kinetic energy. However, even for powerful WR-type winds, the approximation, $L(r) \approx L_\star$, is justified, as the deviations typically remain below $\sim 5\%$ \citep[e.g.][]{Sander2020a}.

Figure~\ref{fig: rad_accl} illustrates the radial variation of total $\Gamma_\mathrm{rad}$ for a representative very massive star model with $M_\star = 105\,M_\odot$ and $T_\star = 35\,\mathrm{kK}$. We decompose this into contributions from bound-bound line transitions, $\Gamma_\mathrm{lines}$, and continuum processes, $\Gamma_\mathrm{cont}$, which includes free-free (ff) and bound-free (bf) continuum terms plus classical electron (Thomson) scattering $\Gamma_\mathrm{e}$:
\begin{equation}
\begin{aligned}
\Gamma_\mathrm{rad} &= \Gamma_\mathrm{lines} + \Gamma_\mathrm{cont} \\
                     &= \Gamma_\mathrm{lines} + \Gamma_\mathrm{ff} + \Gamma_\mathrm{bf} + \Gamma_\mathrm{e}.
\end{aligned}
\label{eq: eddington_parameter_contribution}
\end{equation}
The classical $\Gamma_\mathrm{e}$ provides a useful measure of the star's proximity to the Eddington limit, maintaining near-constancy throughout the atmosphere (solid magenta in Fig.~\ref{fig: rad_accl}). The classical $\Gamma_\mathrm{e}$ is given by:
\begin{equation}
\Gamma_\mathrm{e} = \dfrac{a_\mathrm{thom}}{GM_\star/r^2} \approx  \dfrac{\sigma_\mathrm{e}L_\star}{4\pi GcM_\star},
\label{eq: eddington_parameter_classical}
\end{equation}
where $\sigma_\mathrm{e}$ denotes the electron scattering opacity. For sufficiently high temperatures capable of fully ionising H and He (electron temperatures above approximately $10$ and $50\,\mathrm{kK}$, respectively), the electron scattering opacity remains essentially constant. While this generally holds throughout the inner, sub-critical region of the atmosphere where such conditions are realised, the low-temperature sequences in our grid can show a slight decrease in $\Gamma_\mathrm{e}$ near the critical point. Therefore, the $\Gamma_\mathrm{e}$ values reported in the subsequent sections are averages taken within the sub-critical region.

Two important conclusions emerge from examining the continuum component. First, the true continuum contribution from ff+bf transitions remains vanishingly small relative to electron scattering, evident from the minimal difference between $\Gamma_\mathrm{cont}$ and $\Gamma_\mathrm{e}$ terms. Second, although the continuum component dominates total $\Gamma_\mathrm{rad}$ in the sub-critical inner atmosphere -- accounting for over 70\% of the total -- it cannot launch the wind. Wind launching requires $\Gamma_\mathrm{rad}$ to exceed unity. As the figure demonstrates, continuum processes alone provide insufficient acceleration to overcome gravity; line transitions are essential for wind initiation.

The dominant contributors to line acceleration in the sub-critical region where mass-loss rates are established are line opacities of \ion{Fe}{II}--\ion{Fe}{VII} (depending on $T_\star$). Immediately below the critical point, Fe provides the necessary acceleration to increase total $\Gamma_\mathrm{rad}$ to unity. Despite its relatively low abundance compared to other surface metals, the efficiency of Fe wind driving is due to the spectral coincidence between millions of Fe transitions and the ultraviolet peak characteristic of hot stellar photospheres.

Beyond the critical point, line opacities from several additional elements become important, extending beyond the CNO group. Notable contributors to outer-wind driving include \ion{Si}{}, \ion{P}{}, \ion{S}{}, \ion{Cl}{}, and \ion{Ar}{} \citep[see also][]{Vink99, Sander2017}. Contributions from \ion{Si}{}, \ion{P}{}, and \ion{S}{} are significant in our WNh models, as well as in the B hypergiant models applied to $\zeta^1$ Sco by \citet{Bernini-Peron2025} and in late-WNh models from \citet{Lefever2025}. This contrasts with pure-He classical WR winds, where these contributions are negligible \citep{Sander2020a}, meaning the contribution of \ion{Si}{}, \ion{P}{}, and \ion{S}{} is generally important for stars on or rightward (cooler) of the ZAMS. 

Iron's pivotal role in inner-wind driving is further demonstrated by its absolute and relative contributions to $\Gamma_\mathrm{lines}$. Figure~\ref{fig: relative_contribution_total} presents a stacked plot comparing contributions from various components to total $\Gamma_\mathrm{rad}$ and $\Gamma_\mathrm{lines}$ at the critical point as a function of stellar mass for our $T_\star = 35\,\mathrm{kK}$ temperature sequence. In the top sub-panel, as stellar mass decreases, the $\Gamma_\mathrm{e}$ component at the critical point increases monotonically while line transition contributions decrease. The Fe contribution remains roughly constant at approximately 40\% across most of the parameter space before dropping to roughly 20\% at the lowest masses. In the lowest-mass model, despite Fe contributing only 20\% of total $\Gamma_\mathrm{rad}$, wind launching remains almost entirely Fe-driven. In the bottom sub-panel, Fe contributes more than half of the total line acceleration and remains approximately constant with mass. As we demonstrate in Sect.~\ref{sec: bistability}, insufficient Fe driving can lead to steep drops in the mass-loss rate in certain temperature and mass regimes.

\section{Winds of VMSs}
\label{sec: results}

\subsection{Testing against the transition mass-loss rate}
\label{sec: arches_compare}

\begin{figure*}
    \includegraphics[width = \textwidth]{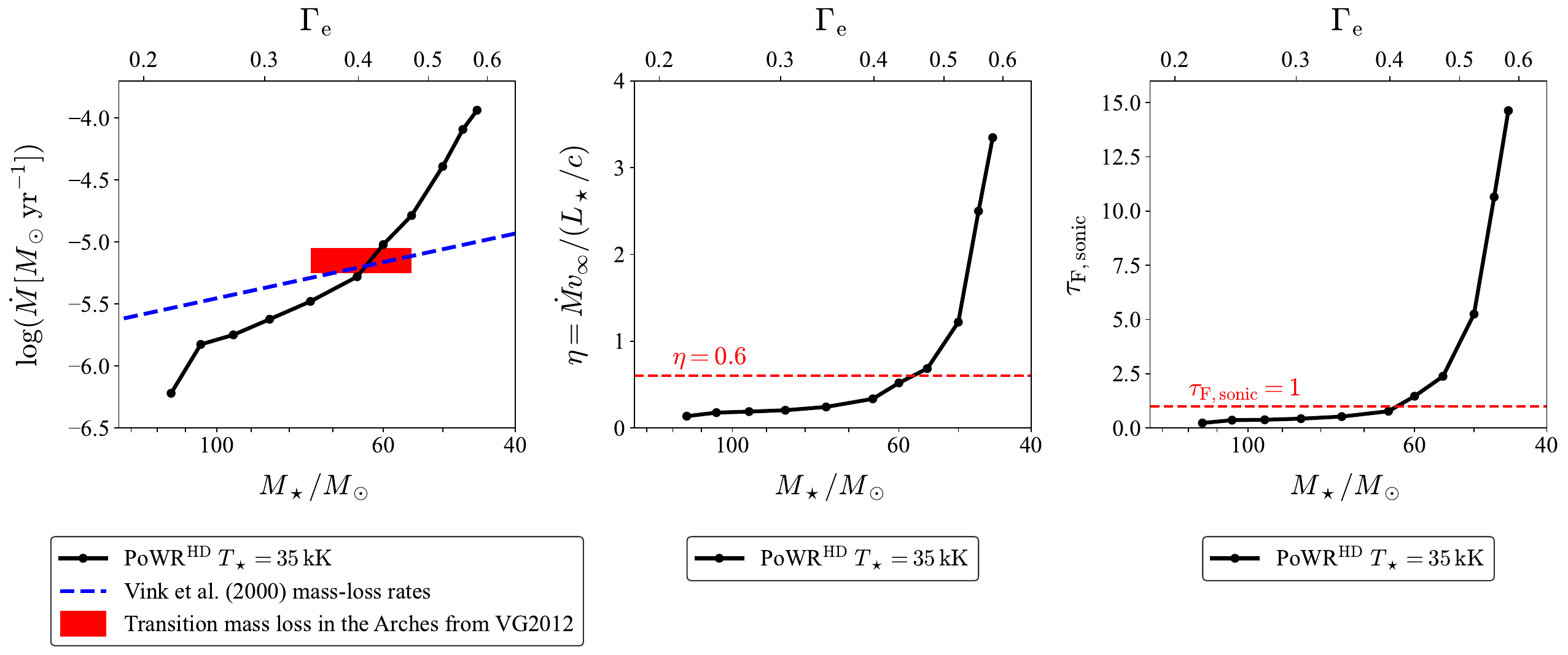}
    \caption{Testing the $\texttt{PoWR}^\textsc{hd}$ model predictions against the transition mass-loss rate in the Arches Cluster.
    \textit{Left:} Predicted mass-loss rates versus stellar mass from $\texttt{PoWR}^\textsc{hd}$ models for the $T_\star = 35\,\mathrm{kK}$ sequence (solid black) and from \citet{Vink2000} (dotted blue), compared to the transition mass-loss rate of $\log(\dot{M}) \approx -5.16$ in the Arches (horizontal red line) from \citet{Vink2012}. The $\texttt{PoWR}^\textsc{hd}$ models are computed with fixed $\log(L_\star/L_\odot) = 6.0$, $X = 0.7$, and $Z = 0.02$, roughly matching the observed properties of the Arches transition objects \citep{Martins2008}. The $T_\star = 35\,\mathrm{kK}$ sequence shown has surface temperatures close to the observed $T_\mathrm{eff}(\tau_\mathrm{R} = 2/3) \approx 33.4\,\mathrm{kK}$ of the transition objects. The stellar masses of the transition objects span $\sim 55-75\,M_\odot$ corresponding to their averaged luminosity and temperature.
    \textit{Middle:} Predicted wind efficiency parameter versus mass compared to the transition value $\eta \approx 0.6$ (dashed red) from \citet{Vink2012}.
    \textit{Right:} Predicted wind optical depth versus mass compared to the transition value $\tau_{F}(r_\mathrm{sonic}) \approx 1$ (dashed red) from \citet{Vink2012}.}
    \label{fig: transition_mdot_Arches}
\end{figure*}

\begin{figure*}
    \includegraphics[width = \textwidth]{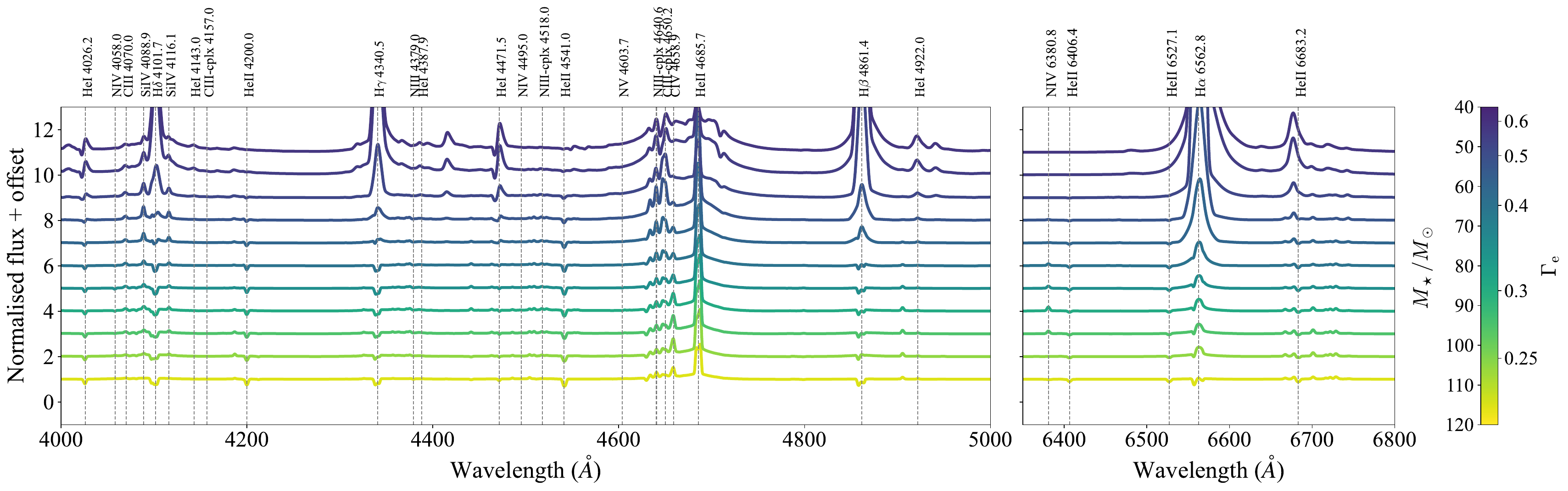}
    \caption{Synthetic spectra showing relevant wind lines in the optical for the $T_\star=35\,\mathrm{kK}$ model sequence presented in Fig.~\ref{fig: transition_mdot_Arches}. The spectra are colour-coded based on the stellar mass. }
    \label{fig: spectra_OPT}
\end{figure*}

\begin{figure*}
    \centering
    \includegraphics[width=\textwidth]{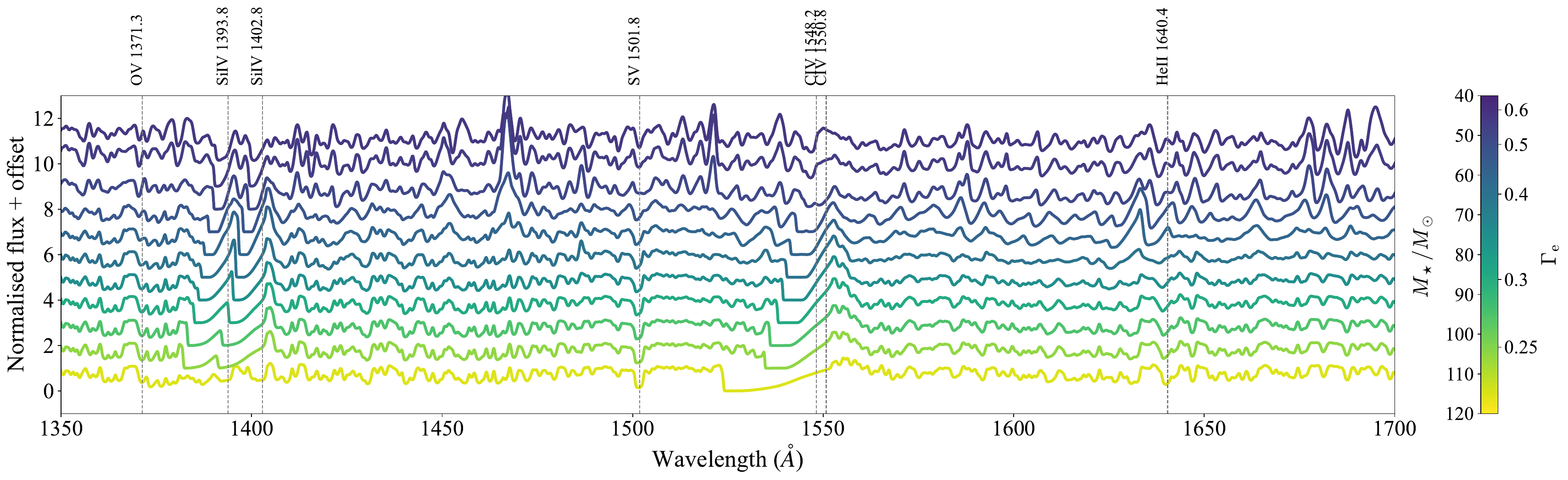}
    \caption{Synthetic spectra showing relevant wind lines in the UV for the $T_\star=35\,\mathrm{kK}$ model sequence presented in Fig.~\ref{fig: transition_mdot_Arches}. The spectra are colour-coded based on the stellar mass.}
    \label{fig: spectra_UV}
\end{figure*}

Determining mass-loss rates for hot, massive stars presents significant challenges, with several complementary approaches available \citep[see][for a recent review]{VinkR}. Empirical determinations typically involve fitting observed wind-sensitive spectral lines such as H$\alpha$ or UV diagnostics using non-LTE expanding atmosphere models. In this iterative process, stellar and wind parameters are varied until optimal spectral fits are achieved, from which one can infer the best-fit empirical mass-loss rates. Alternatively, we can predict theoretical mass-loss rates using analytical models based on CAK theory and its subsequent improvements accounting for complex line-lists and finite-disc effects \citep{CAK1975, FA1986, Pauldrach1986}, semi-analytical approaches such as MC simulations \citep[e.g.][]{LA_1993, Vink2001, MV2008}, or hydrodynamically consistent stellar wind-atmosphere models \citep{GH2005, Sander2020b}, such as those investigated here.

Each approach comes with their own inherent limitations. Optical wind-line diagnostics -- mainly recombination lines -- are inherently model-dependent and strongly sensitive to assumptions about wind clumping in the underlying atmosphere models. As the adopted clumping factor increases, empirically derived rates systematically decrease \citep{Hamann1998}. Furthermore, degeneracies between stellar and wind parameters can make it impossible to uniquely determine best-fit values. Theoretical predictions also suffer from clumping uncertainties, with predicted mass-loss rates depending on clumping assumptions near the critical point \citep{GH2005, Muijres2011, Sabhahit_VMS2025, Bernini-Peron2025}.

To address these challenges, \citet{Vink2012} developed the concept of the transition mass-loss rate. The underlying idea is straightforward: at the observed spectral morphological transition from absorption-dominated O star spectra to emission-line-dominated WNh star winds \citep[as seen in the H$\beta$ transition described by][]{CW2011}, the wind optical depth crosses approximately unity.

Building on the hydrodynamic equation of motion, \citet{Vink2012} derived an analytical formula linking wind efficiency and optical depth to describe this spectral transition condition:
\begin{equation}
\begin{array}{c@{\qquad}c}
\eta \approx f\cdot\tau_{F}(r_\mathrm{sonic}),\;\;
\tau_{F}(r_\mathrm{sonic}) \approx 1.
\end{array}
\label{eq: eta_tau_conditon}
\end{equation}
Through extensive testing across a range of stellar parameters, \citet{Vink2012} found that the factor $f$ depends primarily on the terminal-to-escape velocity ratio $\varv_\infty/\varv_\mathrm{esc}$ and on the velocity-law exponent $\beta$. For Galactic metallicity and $\varv_\infty/\varv_\mathrm{esc} = 2.5$, they obtained $f \approx 0.6$, assuming a $\beta$-type velocity law with relatively large values of $\beta\approx 1.5$, typical of very massive-star winds \citep{Vink2011, Sabhahit_VMS2025}. The dependence of $f$ on the $\varv_\infty/\varv_\mathrm{esc}$ ratio was parametrised in \citet{Sabhahit2023} as $f \sim 0.75\cdot(1+\varv_\mathrm{esc}^2/\varv_\infty^2)^{-1}$\citep[for an analytical derivation, see][]{Grafener2017}. Although clumping assumptions can in principle influence the $\varv_\infty/\varv_\mathrm{esc}$ ratio, and therefore the value of $f$, the impact is small because $f$ approaches a plateau once the $\varv_\infty/\varv_\mathrm{esc}$ ratio is sufficiently high. 

This means the transition from optically thin O-star winds to optically thick WNh-star winds approximately coincides with both the wind efficiency parameter $\eta$ and $\tau_{F}(r_\mathrm{sonic})$ crossing order unity. Beyond this transition, \citet{Vink2011} discovered a steep upturn in predicted mass loss with the classical Eddington parameter, creating the characteristic kink in the $\dot{M}(\Gamma_\mathrm{e})$ relation discussed in the introduction.

\citet{Vink2012}
rearranged Eq.~\eqref{eq: eta_tau_conditon} to formulate the concept of the transition mass-loss rate:
\begin{equation}
\begin{array}{c@{\qquad}c}
\dot{M}_\mathrm{trans} = f \cdot \dfrac{L_\star}{c\varv_\infty}.
\end{array}
\label{eq: mdot_trans}
\end{equation}
As Eq.~\eqref{eq: mdot_trans} reveals, the transition mass-loss rate is essentially a luminosity estimate expressed in mass-loss units. The luminosity and terminal velocity of transition objects can be determined with greater accuracy than empirical measures of the mass-loss rate as the latter depend on model specifics such as clumping assumptions. Uncertainties in $f$ typically fall within the error bars of luminosity measurements. This gives the transition mass-loss rate a crucial advantage: it is both accurate and significantly less model-dependent than traditional estimates. However, this accuracy applies specifically to the transition region between O and WNh stars, not beyond it.

Consider the Arches Cluster which hosts several O stars and WNh stars. The transition objects -- whose spectral morphology bridges absorption-dominated O stars and emission-line-dominated WNh stars -- have luminosities and terminal velocities of approximately $\log(L_\star/L_\odot) \approx 6.06(\pm 0.1)$ and $\varv_\infty \approx 2000(\pm 400)\,\mathrm{km\,s^{-1}}$ \citep{Martins2008}. Applying Eq.~\eqref{eq: mdot_trans} yields a transition mass-loss rate of $\log(\dot{M}_\mathrm{trans}) \approx -5.16$ \citep{Vink2012, Sabhahit2022}.

Given the accurate and model-independent nature of this approach, any credible mass-loss prescription -- whether empirical or theoretical -- should reproduce this rate at the luminosity and temperature of the transition stars. While agreement at the transition does not guarantee accuracy elsewhere, it provides an essential first-order sanity check: if a prescription fails to match the transition mass-loss rate, we can confidently consider it inaccurate within this regime.

Figure~\ref{fig: transition_mdot_Arches} shows how our $\texttt{PoWR}^\textsc{hd}$ models perform against the Arches transition mass-loss rate. We compare our predicted rates for the $T_\star = 35\,\mathrm{kK}$ model sequence with the transition mass-loss rate of $\log(\dot{M}_\mathrm{trans}) \approx -5.16$. The $T_\star = 35\,\mathrm{kK}$ sequence is selected such that its surface temperatures, defined at Rosseland optical depth $\tau_\mathrm{R} = 2/3$ as $T_\mathrm{eff}(\tau_\mathrm{R} = 2/3)$, is close to the observed $T_\mathrm{eff} \approx 33.4\,\mathrm{kK}$ of the transition objects. Our $\texttt{PoWR}^\textsc{hd}$ models successfully reproduce the transition mass-loss rate at stellar parameters corresponding to the Arches transition objects.

We also compare the Arches transition mass-loss rate with the widely used \citet{Vink2000} mass-loss predictions. As shown in Figure~\ref{fig: transition_mdot_Arches} -- and as already noted by \citet{Vink2012} -- the \citet{Vink2000} rates agree well with the transition mass-loss rates for the relevant stellar parameters. However, the \citet{Vink2000} rates diverge from our predictions both above and below the transition point. This deviation becomes particularly pronounced at lower masses or higher $\Gamma_\mathrm{e}$, where our predicted mass loss exhibits a steeper scaling with mass and consequently with $\Gamma_\mathrm{e}$ above the transition. The presence of a kink in the mass-loss rate agrees with the results of \citet{Vink2011}.

Our models predict a terminal velocity of $1500\,\mathrm{km\,s^{-1}}$ at the transition, approximately $500\,\mathrm{km\,s^{-1}}$ lower than observed in Arches transition objects. However, terminal velocity predictions are notoriously sensitive to outer wind conditions, including temperature stratification and clumping, both of which remain uncertain. We can achieve better agreement by increasing outer-wind clumping, which enhances outer-wind driving and boosts terminal velocity. Indeed, such an increase in outer-wind clumping was required to reproduce the UV+optical spectra of R136a1 and R144 using hydrodynamical models \citep{Sabhahit_VMS2025}.

Figure~\ref{fig: transition_mdot_Arches} also compares our predicted wind efficiency parameter and optical depth with transition values from \citet{Vink2012}. Additionally, Figs.~\ref{fig: spectra_OPT} and \ref{fig: spectra_UV} display synthetic optical and UV spectra for the $T_\star=35\,\mathrm{kK}$ model sequence, highlighting crucial wind-sensitive recombination and P-Cygni lines. At the point where our models match the Arches transition mass-loss rate, we find $\eta \approx 0.45$. This value falls below the \citet{Vink2012} transition value of $\eta \approx 0.6$, likely reflecting our underestimated terminal velocities. The wind optical depth crosses unity precisely at the transition, coinciding with the spectral morphological change where H$\beta$ shifts from absorption to emission. We observe similar behaviour in H$\gamma$, while H$\alpha$ transitions from absorption to emission at higher masses (i.e, lower $\Gamma_\text{e}$) compared to other Balmer lines. In contrast, \ion{He}{ii} $\lambda 4686$ remains in emission throughout the sequence, strengthening as mass decreases. In the UV, one can see the expected sensitivity of the the \ion{C}{iv} $\lambda 1550$ and \ion{Si}{iv} $\lambda 1400$ resonance doublets to the terminal velocity in our model. Additional features also appear in the UV, for example around $\lambda 1470$, which is an iron feature that strengthens as the winds become optically thick.

Finally, we can estimate the stellar mass required in our hydrodynamic models to match the transition mass-loss rate. We intend to closely match the (averaged) observed stellar luminosity and surface temperature of the transition object in the Arches. To this end, we interpolate between our model sequences to precisely match the observed $T_\mathrm{eff}(\tau_\mathrm{R} = 2/3) \approx 33.4\,\mathrm{kK}$ \citep{Martins2008}. We find that $M_\star \approx 60\,M_\odot$ (corresponding to $\Gamma_\mathrm{e} \approx 0.43$) and $T_\star \approx 38.2\,\mathrm{kK}$ is required to match the transition. This $M_\star \approx 60\,M_\odot$ coincides with the kink location in our models and agrees well with predictions from \citet{Sabhahit2022} for the location of the kink in $\Gamma_\mathrm{e}$-space.

To summarise our findings: our $\texttt{PoWR}^\textsc{hd}$ models successfully reproduce the transition mass-loss rate in the Arches Cluster, confirm the prediction of a mass-loss kink above the transition, and capture the morphological change in H$\beta$ from absorption to emission as the wind optical depth crosses unity. While our model-predicted wind efficiency at the transition ($\eta \approx 0.45$) is lower than \citet{Vink2012} predicted $\eta \approx 0.6$, this discrepancy likely stems from our underestimated terminal velocities. Overall, these results provide strong validation of our hydro-predicted mass-loss rate and its ability to capture the essential physics of the O to WNh star morphology transition.

\subsection{Wind properties: $\dot{M}, \varv_\infty, \eta$, and $\tau_{F}(r_\mathrm{sonic})$}
\label{sec: wind_properties}

\begin{figure}
    \includegraphics[width = \columnwidth]{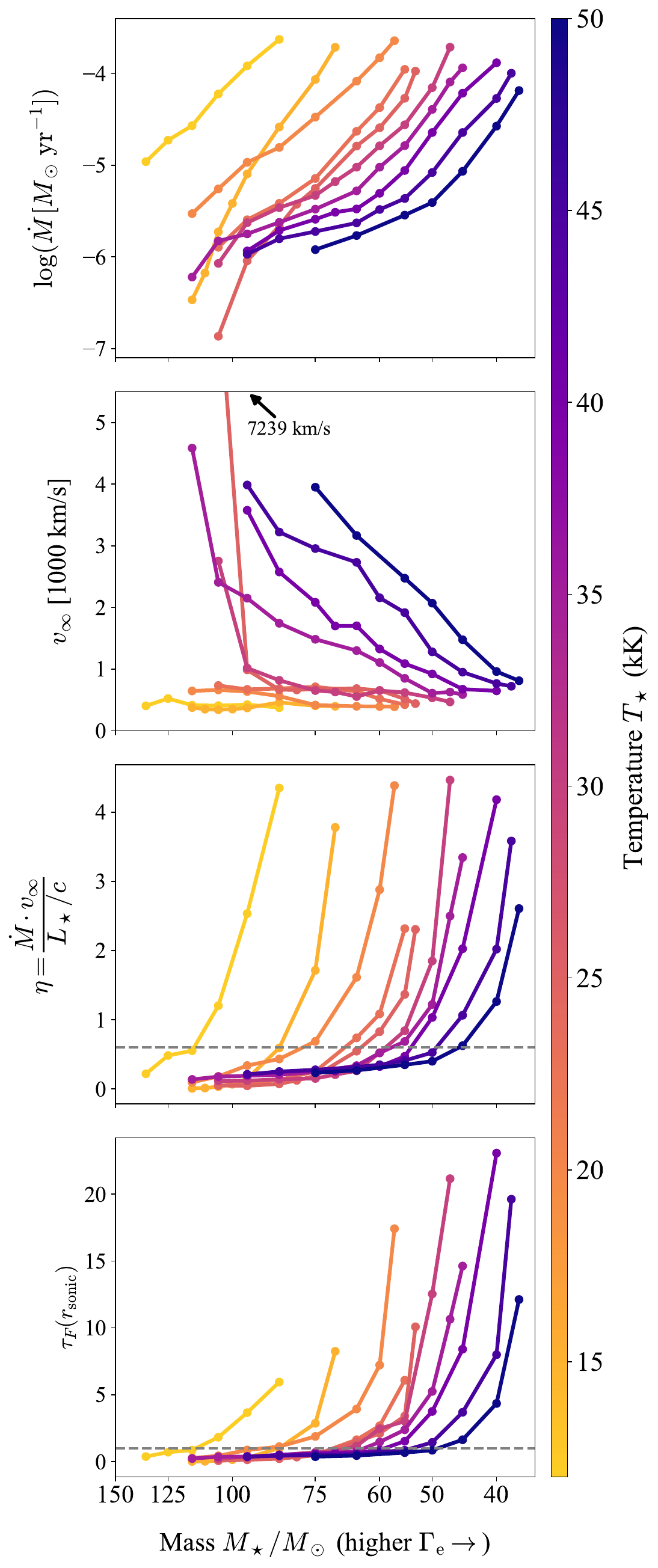}
    \caption{Mass-loss rates, terminal wind velocities, wind efficiencies, and wind optical depths predicted from hydrodynamically consistent $\texttt{PoWR}^\textsc{hd}$ models as a function of stellar mass $M_\star$. The plots are colour-coded according to the different input inner boundary temperatures, ranging from $T_\star = 50$ to $12\,\mathrm{kK}$.  }
    \label{fig: mdot_mass}
\end{figure}

Figure~\ref{fig: mdot_mass} presents relevant wind parameters, including mass-loss rates and terminal wind velocities, as functions of stellar mass, with temperature sequences colour-coded by $T_\star$. Across all temperature sequences, we observe a consistent fundamental trend: mass-loss rates increase systematically towards lower stellar masses when other parameters remain fixed. This behaviour reflects the underlying physics: For fixed luminosity and composition, decreasing the stellar mass decreases the gravitational force in the wind and drives up the total $\Gamma_\mathrm{rad}$ in the sub-critical region ($r < R_\mathrm{crit}$), directly boosting mass-loss rates \citep[cf.][]{CAK1975, LC1999, Vink2001, Sander2020b, Sabhahit_VMS2025}. 

Terminal velocities follow a complementary pattern, generally decreasing with decreasing mass \citep[cf.][]{CAK1975, FA1986}, although the behaviour can be non-monotonic sometimes \citep[see  e.g. the sequence in][]{Lefever2025, Bernini-Peron2025}.  To first order, an increase in the total $\Gamma_\mathrm{rad}$ in the sub-critical regime ($r < R_\mathrm{crit}$) drives a higher mass loss but produces denser, slower winds. Conversely, increased $\Gamma_\mathrm{rad}$ in the supercritical region ($r > R_\mathrm{crit}$) accelerates winds to higher velocities without affecting mass-loss rates \citep[see also][]{ Vink2001, Sander2020b}.

More intriguing are the temperature-dependent variations in these scaling relations. Hotter models with $T_\star \gtrsim 30\,\mathrm{kK}$ reveal a distinctive upturn -- a characteristic kink -- in the $\log\dot{M}-\log M$ (and therefore also the $\log\dot{M}-\log\Gamma_\mathrm{e}$) relation. The scaling behaviour transitions from the shallow slope of $\sim 1.5-3$ at higher masses (lower $\Gamma_\mathrm{e}$) to a dramatically steeper slope of $\sim 10$  at lower masses (higher $\Gamma_\mathrm{e}$). This behaviour qualitatively matches the MC predictions of \citet{Vink2011} and aligns with empirical evidence of such a kink in the 30 Doradus cluster reported by \citet{Best2014}.

In contrast, for the cooler models with $T_\star \lesssim 25\,\mathrm{kK}$, their $\log\,\dot{M}-\log\,M$ relations maintain consistently steep slopes throughout our parameter range, showing no pronounced kink behaviour. Particularly notable are the $T_\star = 25$ and $15\,\mathrm{kK}$ sequences, which maintain distinctly steep slopes across the entire mass range.

Terminal velocity behaviour further distinguishes hot and cool regimes. Hotter models show terminal velocities that generally decline with decreasing mass, while cooler models exhibit weak mass dependence, remaining nearly constant at $\sim 500\,\mathrm{km\,s^{-1}}$. Similarly low velocities were also obtained in the late WNh models of \citet{Lefever2025}.

Both wind efficiency parameters and optical depths increase monotonically with decreasing mass across all temperatures. At kink locations in hotter models, wind efficiency reaches approximately $0.4-0.45$ -- notably lower than the \citet{Vink2012} transition value of $\eta \approx 0.6$, likely reflecting our underestimated terminal velocities. On the other hand, wind optical depths cross unity precisely at these kink locations, supporting our physical interpretation.

\subsection{Role of iron and the bistability jump}
\label{sec: bistability}

\begin{figure}
    \includegraphics[width = \columnwidth]{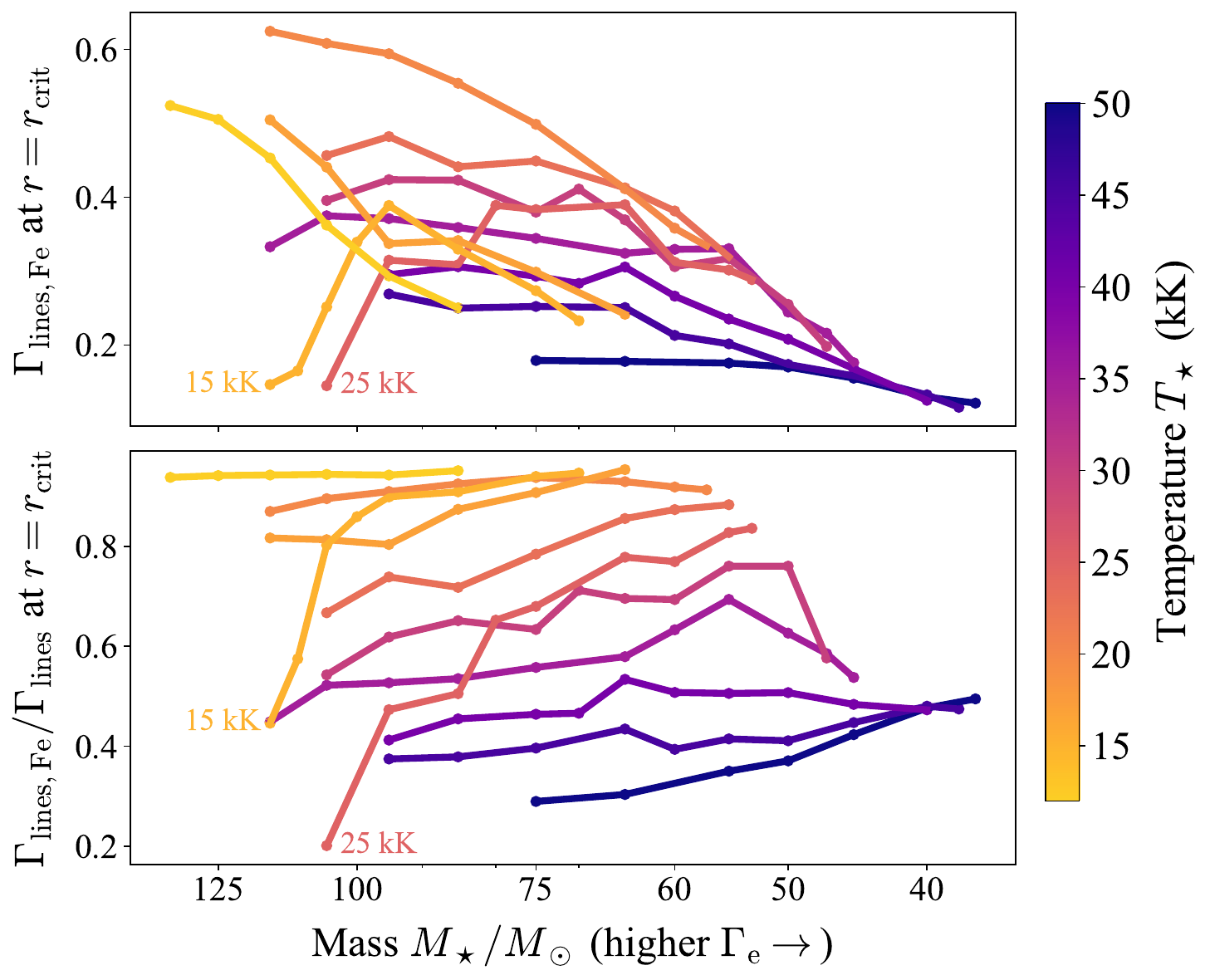}
    \caption{Contribution of Fe-line acceleration at the critical point as a function of mass for different temperature sequences. \textit{Top:} Absolute contribution of Fe. \textit{Bottom:} Relative contribution of Fe to the total line acceleration. The contributions are normalised to gravity and expressed as Eddington parameters. The figure is colour-coded based on the temperature $T_\star$.}
    \label{fig: iron_contribution}
\end{figure}

The steep mass-loss scaling at $T_{\star} = 25$ and $15\,\mathrm{kK}$ demands a deeper investigation into the physical driving mechanisms. Figure~\ref{fig: iron_contribution} demonstrates the fundamental importance of Fe-line acceleration at the critical point. The upper sub-panel reveals a clear trend:  absolute Fe acceleration decreases with mass, consistent with the increasing dominance of $\Gamma_{\mathrm{e}}$ over $\Gamma_{\mathrm{lines}}$ towards lower masses (see Sect.~\ref{sec: rad_acceleration}).

Temperature sequences at $T_{\star} = 25$ and $15\,\mathrm{kK}$, however, stand out with Fe-line contributions declining even at higher masses. While subtle indications of high-mass decline appear at various other temperatures, the effect becomes particularly pronounced at $T_{\star} = 25$ and $15\,\mathrm{kK}$. This behaviour is even more striking in the relative Fe-line contributions to the total $\Gamma_\mathrm{lines}$ (lower sub-panel). At these specific temperatures, the relative Fe contribution drops toward higher masses, contrasting with the approximately constant or weakly declining trends observed at other temperatures. Although Fe continues to dominate sub-critical line acceleration across all models presented, an abrupt reduction occurs precisely at the critical point of outflow initiation for the $T_{\star} = 25$ and $15\,\mathrm{kK}$ sequences. This sudden Fe-driving deficit produces the steep mass-loss rate decline at these characteristic temperatures. This effect was also seen in some of the late-type WNh sequences in \citet{Lefever2025}.

\begin{figure}
    \includegraphics[width = \columnwidth]{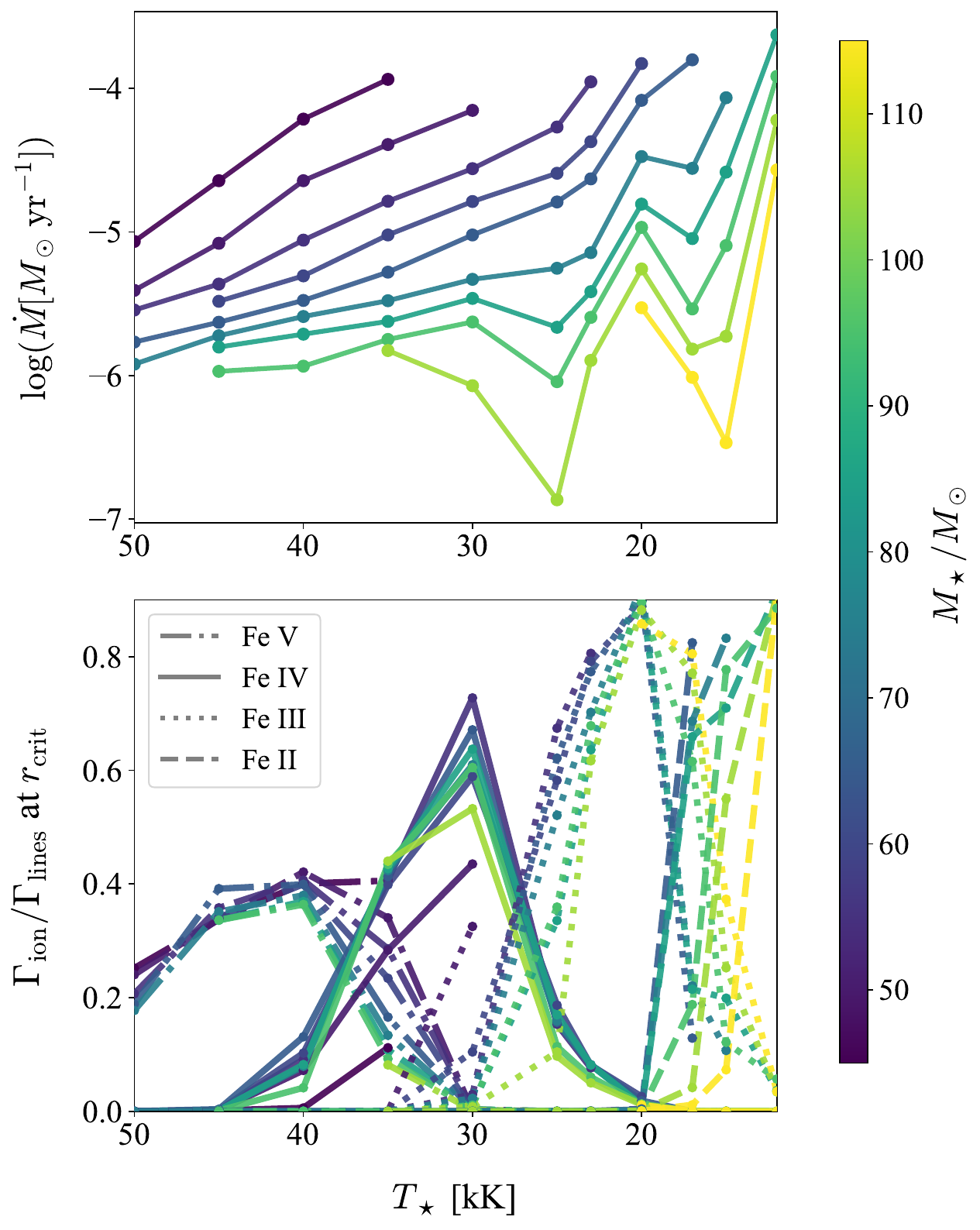}
    \caption{Mass-loss rate as a function of stellar temperature, $T_\star$, for different masses. \textit{(Top:)} Predicted mass-loss rates. \textit{(Bottom:)} Relative contributions from relevant Fe wind-driving ions to the total line acceleration at the critical point. The radiative acceleration contributions are normalised to gravity and expressed as Eddington parameters. The curves are colour-coded by stellar mass.}
    \label{fig: bistability}
\end{figure}

The correspondence between the lack of Fe-line driving and reduced mass-loss rates at temperatures consistent with \citet{Vink99} -- where OB-star wind bistability occurs -- is not coincidental.  The top sub-panel in Figure~\ref{fig: bistability} presents mass-loss rates as functions of $T_{\star}$ for various stellar masses. Stars close to their Eddington limit (corresponding to low mass for a fixed luminosity) exhibit straightforward monotonic behaviour: mass-loss rates increase with decreasing $T_{\star}$, reflecting pure geometrical effect from the increase in radius, that is, for a fixed luminosity, lower temperatures mean larger radii and shallower gravitational potential wells \citep[cf.][]{Sander2023, Lefever2025}. The low-temperature, high $\Gamma_\mathrm{e}$ regime proves particularly relevant for massive LBVs, which exhibit strong winds and, at times, eruptive mass-loss episodes. These models approaching their Eddington limit naturally predict strong winds in the temperature range characteristic of LBVs, consistent with empirical mass-loss rates determined during their quiescent phases.

Models with higher masses, that is, models with lower $\Gamma_\text{e}$, conversely, display complex non-monotonic temperature scaling with sudden jumps in the mass loss below $T_{\star} \approx 25$ and $15\,\mathrm{kK}$. The bottom sub-panel in Figure~\ref{fig: bistability} illustrate the contributions of \ion{Fe}{V}, \ion{Fe}{IV}, \ion{Fe}{III}, and \ion{Fe}{II} to the critical-point line driving. The ionisation sequence evolves predictably with temperature: \ion{Fe}{IV} dominates wind driving at $T_{\star} \approx 30\,\mathrm{kK}$.  As temperatures decreases, \ion{Fe}{IV} recombines into \ion{Fe}{III} which becomes the primary contributor near $T_{\star} \approx 20\,\mathrm{kK}$. At intermediate temperatures around $T_{\star} \approx 25\,\mathrm{kK}$, neither \ion{Fe}{IV} nor \ion{Fe}{III} dominates. Below $T_\star \approx 25\,\mathrm{kK}$, \ion{Fe}{III} begins to dominate the wind driving and we see a rapid increase in the mass-loss rate. 

These results show excellent qualitative agreement with the first bistability jump predicted for B-star winds using MC wind models, both with global and local consistency \citep{Vink99, Vink2018, VS2021}, as well as other wind codes such as CMFGEN and METUJE \citep{Petrov2016, Krticka21}. Empirical evidence appears to exist for the presence of the bistability jump in individual LBV objects \citep[e.g.][]{Groh2009}; but such evidence does not appear to be present in empirical modelling results of O and  B supergiant groups \citep{Crowther2006}. 

The bistability pattern repeats at lower temperatures: \ion{Fe}{III} recombines below $T_\star \approx 20\,\mathrm{kK}$, creating another driving minimum near $T_\star \approx 15\,\mathrm{kK}$ where neither \ion{Fe}{III} nor \ion{Fe}{II} dominates. This produces our second mass-loss rate minimum. Finally, \ion{Fe}{II} takes over below $15\,\mathrm{kK}$, giving rise to yet another bistability jump. Crucially, the Fe-driving minima at $T_\star \approx 25$ and $15\,\mathrm{kK}$ coincide with our models maintaining a steep $\log \dot{M}-\log M$ scaling across the entire investigated mass range (see top sub-panel in $\mathrm{Fig.}\,\ref{fig: mdot_mass}$).

These results demonstrate how ionisation physics governs wind behaviour: the transitions between dominant Fe ionisation states create temperature windows where Fe driving becomes ineffective, which ultimately gives rise to a steep decline in the mass-loss rate in this regime.

\subsection{Mass-loss scaling with mass and temperature}
\label{sec: mass-loss fits}

In this section, we present a fitting relation to predict the mass-loss rate as a function of $M_\star$ and $T_\star$. Since our investigation varies only these parameters, the following formula does not represent a complete mass-loss prescription. Rather, it is designed to capture the scaling relations and functional form, which will serve as a foundation for future work, as we plan to extend this relation by incorporating other relevant parameters in our grid such as luminosity and surface abundances. 

Our fit combines a double-exponential mass scaling -- capturing both shallow and steep scalings -- with quadratic temperature dependence and Gaussian dips reproducing the bistability jumps. These Gaussian terms capture the mass-loss rate minima near $T_{\star} \approx 25\,\mathrm{kK}$ and $15\,\mathrm{kK}$\footnote{All logarithms in the fitting formula are base 10.}:
\begin{equation}
\begin{split}
\log \dot{M}(M_\star,T_\star) &= a
+ \log \Bigg[
10^{\,f_\mathrm{low}(M_\star)}
+ 10^{\,f_\mathrm{high}(M_\star)}
\Bigg] \\
&\quad + b(M_\star)\,\log \!\left(T_\star/T_\mathrm{ref}\right)  
+ c\,\big[\log \!\left(T_\star/T_\mathrm{ref}\right)\big]^2 \\
&\quad - G_1(M_\star) \, \exp\!\Bigg[ 
- \left(\dfrac{T_\star - T_1}{\sigma_T}\right)^2 \Bigg] \\
&\quad - G_2(M_\star) \, \exp\!\Bigg[  
-\left(\dfrac{T_\star - T_2(M_\star)}{\sigma_T}\right)^2 \Bigg],
\label{eq: mdot_fit}
\end{split}
\end{equation}
where $M_{\star}$ is in solar masses and $T_{\star}$ is in kiloKelvin. The bistability temperatures $T_1$ and $T_2(M_\star)$ mark the centres of the Gaussian dips, while $\sigma_T$ controls their width. We fix $T_1 = 25\,\mathrm{kK}$ and $\sigma_T = 2\,\mathrm{kK}$ during fitting, as our models demonstrate negligible variation of these parameters with mass or temperature. Stellar mass and temperature are normalised to references values, $M_\mathrm{ref}$ and $T_\mathrm{ref}$, with $M_\mathrm{ref}=60\,M_\odot$ at $T_\mathrm{ref} = 38\,\mathrm{kK}$. These normalisation values are not arbitrary -- as demonstrated in Sect.~\ref{sec: arches_compare}, they roughly correspond to the $M_\star$ and $T_\star$ values required by our hydrodynamic models to reproduce the Arches Cluster transition mass-loss rate. The mass-dependent coefficients take the following form:
\[
\begin{aligned}
a &= -5.527\,(\pm0.022) \\
f_\mathrm{low}(M_\star) &= -1.864\,(\pm0.198) \cdot\log(M_\star/M_\mathrm{ref}), \\
f_\mathrm{high}(M_\star) &= -9.865\,(\pm0.273)\cdot\log(M_\star/M_\mathrm{ref}) \\
b(M_\star) &= -2.062\,(\pm0.521) + 5.671\,(\pm1.536) \cdot\log(M_\star/M_\mathrm{ref}) \\
c &= 3.974\,(\pm0.459)  \\
G_1(M_\star) &= 0.178\,(\pm0.042) \;\exp\big[9.611\,(\pm1.293) \cdot\log(M_\star/M_\mathrm{ref})\big] \\
G_2(M_\star) &= 0.291\,(\pm0.051) \;\exp\big[8.658\,(\pm1.048) \cdot \log(M_\star/M_\mathrm{ref})\big] \\
T_2(M_\star) &= 16.941\,(\pm0.268) - 7.274\,(\pm1.506) \cdot\log(M_\star/M_\mathrm{ref}), \\
M_\mathrm{ref} &= 60 - 0.521(\pm0.133) \cdot(T_\star-T_\mathrm{ref}) \\
T_1 &= 25\,\mathrm{kK}, \quad \sigma_T = 2\,\mathrm{kK}, \quad T_\mathrm{ref} = 38\,\mathrm{kK}.
\end{aligned}
\]

\begin{figure}
    \includegraphics[width = \columnwidth]{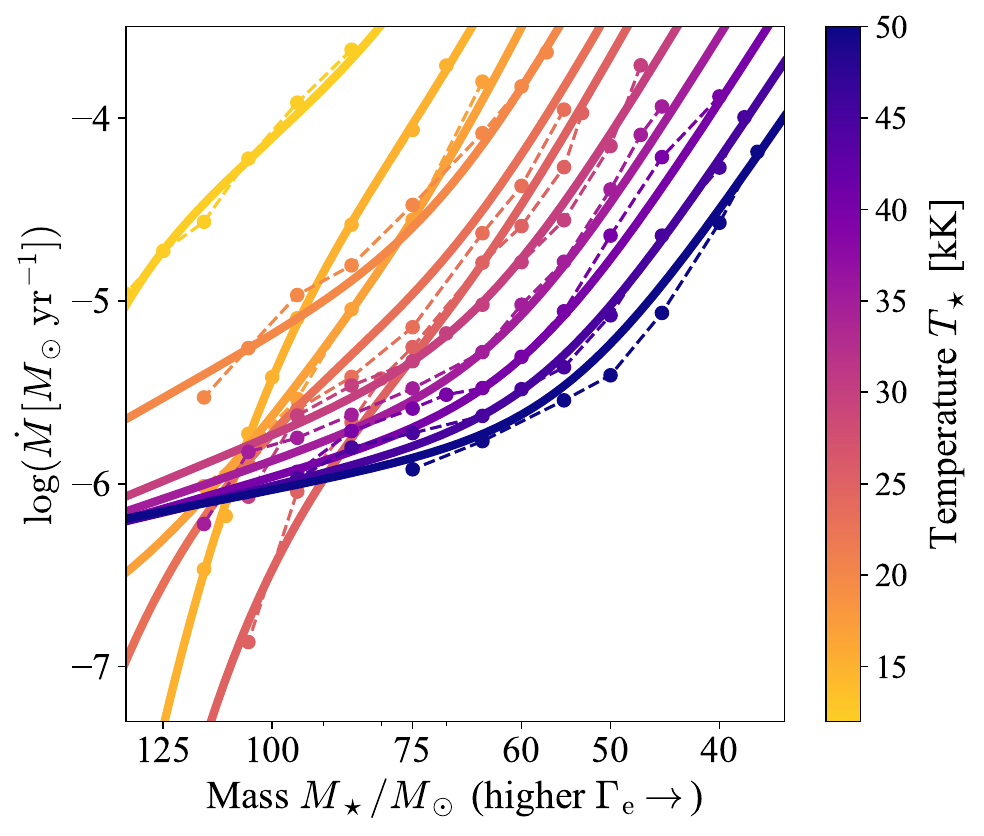}
    \caption{$\texttt{PoWR}^{\textsc{hd}}$ predicted mass-loss rates (symbols with dashed lines) compared to our best-fit relation (solid lines) as functions of stellar mass, $M_{\star}$. The fit derives from Eq.~\eqref{eq: mdot_fit}, with lines colour-coded by stellar temperature, $T_{\star}$.}
    \label{fig: mdot_M_fit}
\end{figure}

Several simplifications help elucidate the different scaling relations in Eq.~\eqref{eq: mdot_fit}. Our fitting function ensures that at high temperatures ($T_\star \gtrsim 30\,\mathrm{kK}$) -- beyond the influence of bistability features -- the Gaussian corrections vanish,  leaving only the baseline mass and temperature scaling. The double-exponential term $10^{f_{\mathrm{high}}(M_{\star})} + 10^{f_{\mathrm{low}}(M_{\star})}$ captures a critical feature: the transition from shallow to steep power-law scaling of $\log \dot{M}$ versus $\log M_\star$ at higher temperatures. For the simplified case of $T_{\star} = 38\,\mathrm{kK}$, Eq.~\eqref{eq: mdot_fit} reduces to 
\begin{equation}
\begin{split}
\log \dot{M}(M_\star,38\,\mathrm{kK}) \approx a + \log \left[ 10^{\,f_\mathrm{low}(M_\star)} + 10^{\,f_\mathrm{high}(M_\star)} \right],
\label{eq: mdot_fit_T_fixed_38}
\end{split}
\end{equation}

which simplifies further depending on which term dominates based on the stellar mass relative to $60\,M_\odot$. For high-mass stars with $M_\star > 60\,M_\odot$ or low $\Gamma_\mathrm{e}$, the first term dominates, while for lower-mass stars, $M_\star < 60\,M_\odot$, with a high $\Gamma_\mathrm{e}$, the second term takes over:

\begin{equation}
\begin{split}
\log \dot{M}(M_\star,38\,\mathrm{kK})\approx \begin{cases}
a - 1.864 \cdot \log(M_\star/60) & \text{if } M_\star > 60 \\
& \text{or low } \Gamma_\mathrm{e} \\
a - 9.865 \cdot \log(M_\star/60) & \text{if } M_\star < 60 \\
& \text{or high } \Gamma_\mathrm{e}.
\end{cases}
\label{eq:mdot_fit_reduced_based_on_mass}
\end{split}
\end{equation}
The double-exponential term therefore accurately captures the shallow-to-steep transition in the $\log\dot{M}$ versus $\log M_\star$ scaling predicted by our models. Figure~\ref{fig: mdot_M_fit} shows our fit relations, where for $T_\star \gtrsim 30\,\mathrm{kK}$, the $\log\,\dot{M}$ versus $\log M_\star$ scaling steepens below $M_\star \approx 60\,M_\odot$.

\begin{figure}
    \includegraphics[width = \columnwidth]{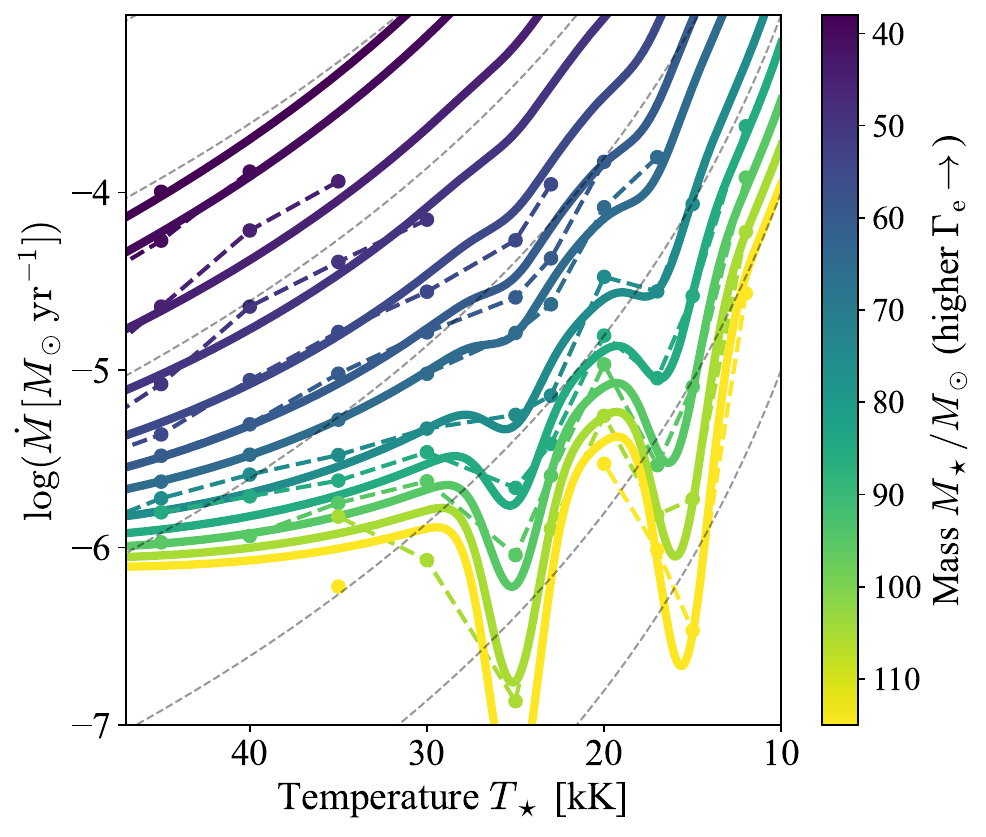}
    \caption{Same as Fig.~\ref{fig: mdot_M_fit}, but presenting mass-loss rates as functions of stellar temperature, $T_{\star}$, colour-coded by stellar mass, $M_{\star}$. The dashed black lines indicate reference scalings of $\dot{M} \propto T_\star^{-6}$ scaling, which arise from geometric effects.}
    \label{fig: mdot_T_fit}
\end{figure}

To understand how the baseline mass-loss rate scales with temperature, we fix $M_\star = 60\,M_\odot$. Neglecting the minor variation of $M_\mathrm{ref}$ with $T_\star$ and the double-Gaussian bistability dips, the fit relation reduces to
\begin{equation}
\begin{split}
\log \dot{M}(60\,M_\odot,T_\star) \approx a + \mathrm{log}2 - 2.062\cdot \mathrm{log}(T_\star/38) \\
\,+\,  3.974\cdot\big[\log \!\left(T_\star/38\right)\big]^2 .
\label{eq: mdot_fit_M_fixed_60}
\end{split}
\end{equation}
In Fig.~\ref{fig: mdot_T_fit}, we show our fitted relation as a function of $T_\star$ from Eq.~\eqref{eq: mdot_fit}, compared to our model predictions. The baseline scaling of $\log\dot{M}(T_\star)$ with $\log T_\star$ exhibits a steep negative dependence. For example, at $T_\star = 38\,\mathrm{kK}$, differentiating Eq.~\eqref{eq: mdot_fit_M_fixed_60} gives an instantaneous slope of $\dot{M} \propto T_\star^{-2.062}$. As $T_\star$ decreases, this slope  further steepens. We also compare our predictions with reference lines showing a $\dot{M} \propto T_\star^{-6}$ scaling, which arise from the geometrical effect of increasing radii and decreasing surface gravity. We find that in certain parameter regimes, the mass-loss scaling with temperature is dominated by this purely geometrical effect, particularly at higher $\Gamma_\mathrm{e}$ and lower $T_\star$.

We acknowledge a practical limitation of Eq.~\eqref{eq: mdot_fit}: the temperature $T_\star$ is defined at an arbitrary Rosseland continuum optical depth of $\tau_\mathrm{R,cont} = 20$, which may not be readily accessible in stellar structure and evolution models \citep[see e.g. the recent discussion in][]{Josiek2025}. Most structure codes compute the surface temperature $T_{\mathrm{eff}}$ at the photospheric surface where $\tau_{\mathrm{R}} = 2/3$, assuming a static, grey atmosphere. Furthermore, due to the arbitrary choice of $\tau_\mathrm{R,cont}$, defining the effective temperature at $\tau_{\mathrm{R}} = 2/3$ provides a more physically meaningful measure of stellar temperature.

\begin{figure}
    \includegraphics[width = \columnwidth]{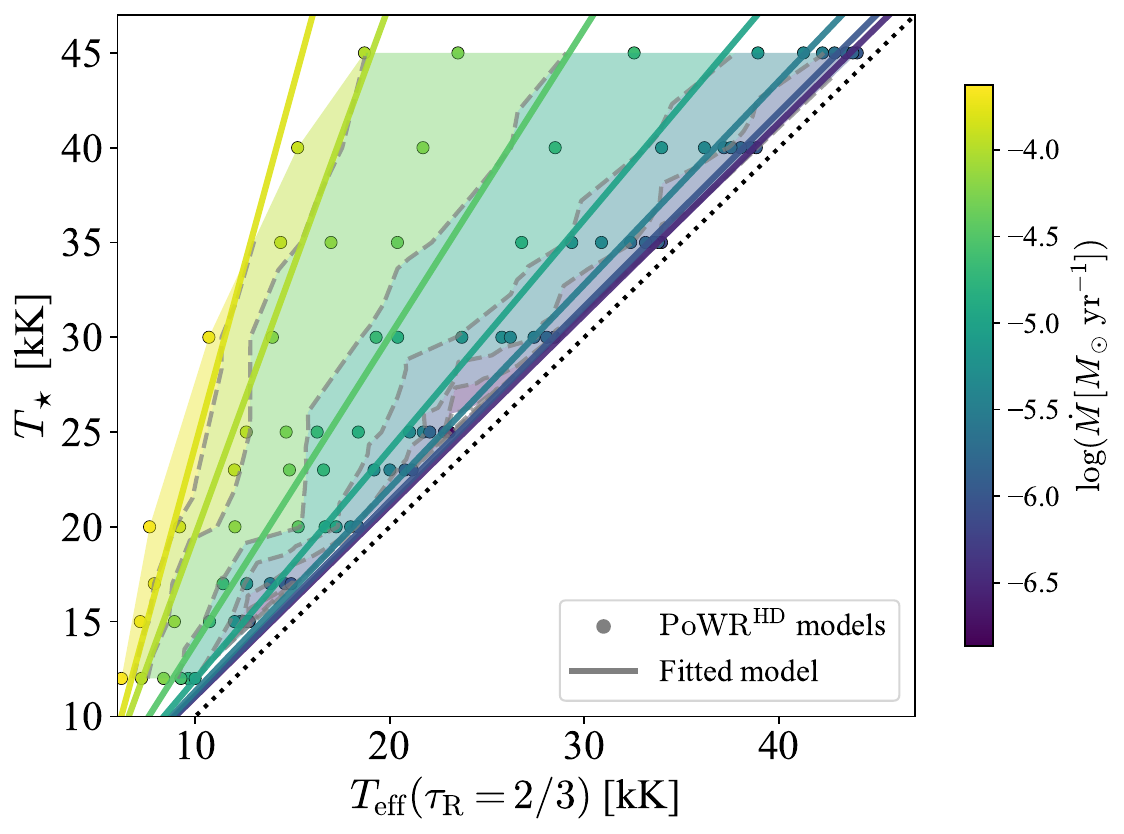}
    \caption{Change in the inner boundary temperature $T_\star$ as a function of the effective temperature defined at Rosseland optical depth $\tau_{\mathrm{R}} = 2/3$ as winds become optically thick. Symbols show our $\texttt{PoWR}^{\textsc{hd}}$ model predictions, colour-coded by mass-loss rate. Contours of constant mass-loss rate are shown with dashed lines, while solid lines denote the fits from Eq.~\eqref{eq: Tstar_T23_Mdot_fit}. The dotted black line marks where the two temperatures are equal.}
    \label{fig: deltaT_mdot_fit}
\end{figure}

We therefore provide a supplementary relation connecting mass-loss rate with both $T_{\star}$ and $T_{\mathrm{eff}}(\tau_{\mathrm{R}} = 2/3)$ for our atmosphere models, enabling temperature parameter mapping based on the predicted $\dot{M}$. Figure~\ref{fig: deltaT_mdot_fit} illustrates the relationship between $T_{\star}$ and $T_{\mathrm{eff}}(\tau_{\mathrm{R}} = 2/3)$. The difference between the two temperatures increases with increasing mass-loss rate \citep[cf.][]{Smith2004, Sander2023}, as the wind becomes more optically thick. We adopt a linear relation between $T_{\star}$ and $T_{\mathrm{eff}}(\tau_{\mathrm{R}} = 2/3)$, with both slope and y-intercept scaling with mass-loss rate:
\begin{equation}
T_\star = a(\dot{M}) + b(\dot{M})\cdot T_{\mathrm{eff}}(\tau_{\mathrm{R}} = 2/3),
\label{eq: Tstar_T23_Mdot_fit}
\end{equation}
where
\begin{equation}
\begin{split}
a(\dot{M}) &= 1 -0.85(\pm 0.277)\;\Bigg(\dfrac{\dot{M}}{\dot{M}_\mathrm{ref}}\Bigg)^{0.901(\pm0.097)} \\
b(\dot{M}) &= 1 + 0.152(\pm0.14)\;\Bigg(\dfrac{\dot{M}}{\dot{M}_\mathrm{ref}}\Bigg)^{0.925(\pm0.319)} \\
\dot{M}_\mathrm{ref} &= 6.92\cdot10^{-6}\;M_\odot\,\mathrm{yr^{-1}}.
\label{eq: Tstar_T23_Mdot_ab_fit}
\end{split}
\end{equation}
Here, $\dot{M}$ is in $M_{\odot}\,\mathrm{yr}^{-1}$, while temperatures are in kK.  The coefficients $a(\dot{M})$ and $b(\dot{M})$ are constructed such that at very low mass loss, the difference saturates to $\Delta T = T_\star - T_{\mathrm{eff}}(\tau_{\mathrm{R}} = 2/3) \sim 1\,\mathrm{kK}$ in agreement with our model predictions in Fig.~\ref{fig: deltaT_mdot_fit}. While the typical error from our $T_\star$ fits is of the order of $1\,\mathrm{kK}$, especially for the highest mass-loss rate models, it successfully captures the overall trend of increasing difference between the temperatures towards higher mass-loss rate. Therefore, given $M_{\star}$ and $T_{\mathrm{eff}}(\tau_{\mathrm{R}} = 2/3)$, an iterative solution of Eqs.~\ref{eq: mdot_fit} and \ref{eq: Tstar_T23_Mdot_fit} simultaneously yields $\dot{M}$ and $T_{\star}$.

While we present fits capturing the temperature mapping between $T_\star$ and $T_{\mathrm{eff}}(\tau_{\mathrm{R}} = 2/3)$ for the current set of atmosphere models, we emphasise that these relations are preliminary and should not be over-interpreted. For example, The $T_{\mathrm{eff}}(\tau_{\mathrm{R}} = 2/3)$ values calculated in evolution models assume a grey static atmosphere connected to the outer boundary. Consequently, the $T_{\mathrm{eff}}(\tau_{\mathrm{R}} = 2/3)$ from evolution models could be inaccurate, due to effects such as inflation and turbulence associated with the hot iron bump \citep[see e.g.][for potential issues with connecting structure and atmosphere models]{Josiek2025}. Additionally, since we varied only $M_\star$ in this work, further testing with variations in other stellar parameters is necessary to establish whether this correction is universally applicable.

Finally, we perform a sanity check of our fitting relations using the transition mass-loss rate obtained from applying Eq.~\eqref{eq: mdot_trans} to the transition objects in the Arches Cluster. Section~\ref{sec: arches_compare} demonstrated that the mass-loss kink occurs at $M_{\star} \approx 60\,M_{\odot}$. With mass dependencies normalised to $60\,M_{\odot}$ and $T_{\mathrm{eff}}(\tau_{\mathrm{R}} = 2/3) = 33.4\,\mathrm{kK}$, solving Eqs.~\ref{eq: mdot_fit} and \ref{eq: Tstar_T23_Mdot_fit} gives $\log \dot{M} \approx -5.22$ (and $T_{\star} = 38.05\,\mathrm{kK}$). This value closely matches the Arches transition mass-loss rate, validating our fitting relations.

\section{Discussion: Comparison to previous works}
\label{sec: discussion}

We now place our model predictions in context with previous VMS wind investigations. \citet{GH2008} examined VMS winds across a comparable parameter space using a precursor to our $\texttt{PoWR}^{\textsc{hd}}$ version. Our models improve upon theirs through expanded atomic data and broader coverage of the $\Gamma_{\mathrm{e}}-T_{\star}$ parameter space (see Appendix~\ref{appendix: summary}). While \citet{GH2008} identified steep scaling of mass loss with $\Gamma_{\mathrm{e}}$, their predicted mass-loss rates lacked kink behaviour -- possibly because their models did not cover the $\Gamma_{\mathrm{e}} < 0.4$ regime where slope shallowing occurs in our calculations. We further check the absolute mass-loss rate predictions for $M_\star=60\,M_\odot$ and $T_\star=38\,\mathrm{kK}$ where we identified our kink to match the transition mass-loss rate. There is roughly 0.2 dex difference between our predictions, with our rates higher compared to \citet{GH2008} (see Fig.~\ref{fig: compare_to_graf}). This difference could be due to our expanded atomic data, as well as our inclusion of turbulent pressure in the atmosphere.

For high-$\Gamma_{\mathrm{e}}$ VMSs, our models predict a steep mass-loss scaling with mass and consequently $\Gamma_{\mathrm{e}}$ of $\dot{M}\sim \Gamma_\mathrm{e}^{10}$ (see Fig.~\ref{fig: mdot_M_fit}). Although further work must establish whether such scaling persists in the $\dot{M}-\Gamma_{\mathrm{e}}$ relation when varying luminosity and H content simultaneously, this exceeds the $\sim 5$ scaling reported by \citet{GH2008} and \citet{Vink2011}. 

The overlapping temperature range ($30-50\,\mathrm{kK}$) between our work and \citet{GH2008} also reveals qualitative agreement between both studies: the baseline mass-loss rate increases with decreasing $T_{\star}$.  

VMS winds were further studied using MC techniques by \citet{Vink2011}, who predicted the kink behaviour. While qualitative agreement exists between our predictions, the location of the kink in our models occurs at $\Gamma_\mathrm{e}$ close to 0.4, lower compared to \citet{Vink2011}. However, this is consistent with \citet{Vink2011} expectations that their kink would shift toward lower values.

\section{Conclusions}
\label{sec: conclusions}

In this paper, we have investigated VMS wind properties using hydrodynamically consistent $\texttt{PoWR}^{\textsc{hd}}$ stellar atmosphere models. We examined how stellar mass and temperature govern predicted mass-loss rates, terminal velocities, and wind characteristics across the parameter space relevant to WNh stars. Our principal findings are s follows:

\begin{enumerate}

    \item Our models predict a characteristic upturn in the $\log \dot{M}-\log M$ relation for sequences with $T_{\star} \gtrsim 30\,\mathrm{kK}$, producing a kink feature consistent with Monte Carlo simulations of \citet{Vink2011} and empirical evidence from \citet{Best2014}. The location of the kink in the MC models roughly coincides with the crossing of the single-scattering limit and the wind optical depth reaching order unity. With multiple scattering in the wind, the star can drive a stronger mass-loss rate, resulting in the appearance of a kink. Our CMF-based models corroborate the MC results; for instance, our models also predict the characteristic kink feature where the wind optical depth crosses unity, and the synthetic spectra transition from absorption-dominated O-type to emission-dominated H-rich WR morphologies. The wind efficiency parameter reaches $\eta \sim 0.45$ at the kink -- slightly below the $\eta \sim 0.6$ predicted by \citet{Vink2012}. This discrepancy likely reflects our under-prediction of terminal velocities compared to those inferred for transition-type stars in the Arches Cluster. For the chosen luminosity, the kink occurs at $M_{\star} \approx 60\,M_{\odot}$ corresponding to $\Gamma_{\mathrm{e}} \approx 0.43$. 

    \item We tested our predictions against the model-independent transition mass-loss rate in the Arches Cluster near the Galactic center, offering a crucial first-order sanity check for any prescribed mass loss at the O to WNh boundary. Our mass-loss rates match the Arches transition value of $\log(\dot{M}_{\mathrm{trans}}) \approx -5.16$ \citep{Vink2012} at our predicted kink location.  Importantly, we also reproduce the associated spectroscopic transition from absorption- to emission-dominated H$\beta$ profiles as the wind optical depth passes unity, thereby confirming the physical interpretation of the transition mass-loss concept and its connection to optically thick wind onset in VMSs.

    \item In addition to the kink, our models reveal dramatic temperature-dependent jumps in the predicted mass-loss rates, most prominently at $T_\star \approx 25$ and $15\,\mathrm{kK}$. These features arise from shifts in the dominant Fe ionisation stage -- \ion{Fe}{iv}$\,\rightarrow\,$\ion{Fe}{iii} near $25\,\mathrm{kK}$ and \ion{Fe}{iii}$\,\rightarrow\,$\ion{Fe}{ii} near $15\,\mathrm{kK}$ --creating radiative driving deficits where both upper and lower ionisation stages dominate. The resulting bistability jumps agree qualitatively with Monte Carlo line-driven wind predictions from both globally and locally consistent models \citep{Vink99, Vink2018, VS2021}, as well as CMFGEN and METUJE computations \citep{Petrov2016, Krticka21}. At these specific temperatures, our predicted $\log\dot{M}-\log M$ relation maintains steep scaling throughout the investigated mass range.

    \item  A radiative acceleration analysis reveals the decisive role of Fe-group elements, which -- despite low abundances -- contribute over 50\% of total line driving through myriad UV transitions. In the sub-critical wind regime where $\dot{M}$ is set, continuum processes alone are insufficient to overcome gravity and Fe line transitions initiate the wind.

    \item To facilitate further applications, we provide fitting prescriptions (Equations~\ref{eq: mdot_fit}, \ref{eq: Tstar_T23_Mdot_fit}) that reproduce both the kink and the bistability behaviour. These relations combine a double-exponential mass scaling with Gaussian terms capturing the ionisation-driven minima, thereby encoding the complex temperature and mass dependencies. Although derived for fixed luminosity and composition, these formulae establish a quantitative foundation for future incorporation of additional parameters including luminosity, surface H fraction, and metallicity.

\end{enumerate}

Together, these results represent significant progress toward predictive VMS mass-loss prescriptions. By capturing both the Eddington-parameter associated kink behaviour and Fe-ionisation shift driven bistable winds, our hydrodynamically consistent models address fundamental gaps in VMS wind theory, with implications for stellar evolution, chemical feedback, and stellar fate.

\begin{acknowledgements}
We thank the anonymous referee for constructive comments that helped improve the paper. GNS and JSV are supported by STFC funding under grant number ST/Y001338/1. AACS is supported by the German
    \emph{Deut\-sche For\-schungs\-ge\-mein\-schaft, DFG\/} in the form of an Emmy Noether Research Group -- Project-ID 445674056 (SA4064/1-1, PI Sander). AACS further acknowledges financial support by the Federal Ministry for Economic Affairs and Climate Action (BMWK) via the German Aerospace Center (Deutsches Zentrum f\"ur Luft- und Raumfahrt, DLR) grant 50 OR 2503 (PI: Sander). This project was co-funded by the European Union (Project 101183150 - OCEANS).
\end{acknowledgements}

\bibliographystyle{aa}
\bibliography{References.bib}

\appendix

\twocolumn[\section{Additional plots}\label{appendix: mdot_solution}]

\begin{figure}[!htb]
    \centering
    \includegraphics[width=\columnwidth]{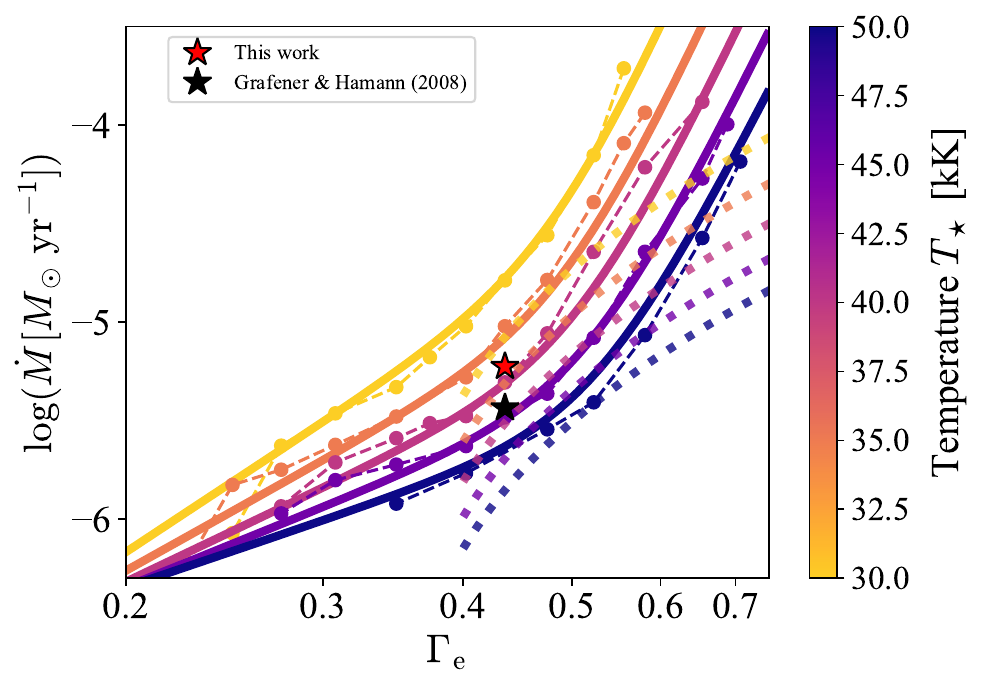}
    \caption{Mass-loss rate versus the classical Eddington parameter $\Gamma_\mathrm{e}$, comparing our predictions with those of \citet{GH2008}. Our model predictions are indicated by coloured symbols connected with dashed lines. The best-fit relation from Eq~\eqref{eq: mdot_fit} is shown as solid lines, while the fit from \citet{GH2008} is shown as dotted lines. Absolute mass-loss rates predicted by the two fits for $M_\star = 60\,M_\odot$ and $T_\star = 38\,\mathrm{kK}$ are indicated by red and black star symbols, respectively.}
    \label{fig: compare_to_graf}
\end{figure}

\section{Parameter space and predicted wind properties}
\label{appendix: summary}

In our $\texttt{PoWR}^\textsc{hd}$ models, we use an extended set of 17 elements with their respective mass fractions provided in Table~\ref{tab: metal_mass_fraction}. The total metal mass fraction adds up to $Z= 0.02$ with individual metal mass fractions distributed following solar-scaled abundances from \citet{GS98}. The detailed breakdown of the relevant ion list and their respective level and line numbers are provided in Table~\ref{tab: line_list}

\begin{table}[H]
\centering
\renewcommand{\arraystretch}{1.15} 
\caption{Mass fractions of elements used in the `base model' (Z = 0.02).}
\begin{tabular}{c c} 
\hline
Element & Mass fraction \\
\hline
$X$ &0.7 \\
$Y$ & $ 0.28$ \\
$Z_\mathrm{C}$  & $3.4416\times10^{-3}$ \\
$Z_\mathrm{N}$  & $1.0082\times10^{-3}$ \\
$Z_\mathrm{O}$  & $9.3605\times10^{-3}$ \\
$Z_\mathrm{Ne}$ & $2.0995\times10^{-3}$ \\
$Z_\mathrm{Na}$ & $4.1565\times10^{-5}$ \\
$Z_\mathrm{Mg}$ & $7.9963\times10^{-4}$ \\
$Z_\mathrm{Al}$ & $7.2154\times10^{-5}$ \\
$Z_\mathrm{Si}$ & $8.8242\times10^{-4}$ \\
$Z_\mathrm{P}$  & $9.7317\times10^{-6}$ \\
$Z_\mathrm{S}$  & $4.3977\times10^{-4}$ \\
$Z_\mathrm{Cl}$ & $5.8458\times10^{-6}$ \\
$Z_\mathrm{Ar}$ & $8.6834\times10^{-5}$ \\
$Z_\mathrm{K}$  & $4.5641\times10^{-6}$ \\
$Z_\mathrm{Ca}$ & $7.7642\times10^{-5}$ \\
$Z_\mathrm{Fe}$ & $1.6649\times10^{-3}$ \\
\hline
\end{tabular}
\label{tab: metal_mass_fraction}
\end{table}

\begin{table}[H]
\centering
\caption{List of ions and their corresponding level and line numbers used in our $\texttt{PoWR}^\textsc{hd}$ models.}
\begin{tabular}{c c c c c c}
\hline
Ion & Levels & Lines & Ion & Levels & Lines \\
\hline
\ion{H}{I} & 30 & 435 & \ion{Si}{V} & 52 & 1326 \\
\ion{H}{II} & 1 & 0 & \ion{Si}{VI} & 1 & 0 \\
\ion{He}{I} & 45 & 990 & \ion{P}{III} & 47 & 1081 \\
\ion{He}{II} & 30 & 435 & \ion{P}{IV} & 12 & 66 \\
\ion{He}{III} & 1 & 0 & \ion{P}{V} & 11 & 55 \\
\ion{C}{I} & 1 & 0 & \ion{P}{VI} & 1 & 0 \\
\ion{C}{II} & 10 & 45 & \ion{S}{III} & 23 & 253 \\
\ion{C}{III} & 40 & 780 & \ion{S}{IV} & 25 & 300 \\
\ion{C}{IV} & 25 & 300 & \ion{S}{V} & 20 & 190 \\
\ion{C}{V} & 29 & 406 & \ion{S}{VI} & 2 & 1 \\
\ion{C}{VI} & 1 & 0 & \ion{Cl}{III} & 1 & 0 \\
\ion{N}{I} & 1 & 0 & \ion{Cl}{IV} & 24 & 276 \\
\ion{N}{II} & 30 & 435 & \ion{Cl}{V} & 18 & 153 \\
\ion{N}{III} & 87 & 3741 & \ion{Cl}{VI} & 1 & 0 \\
\ion{N}{IV} & 38 & 703 & \ion{Ar}{III} & 1 & 0 \\
\ion{N}{V} & 20 & 190 & \ion{Ar}{IV} & 13 & 78 \\
\ion{N}{VI} & 1 & 0 & \ion{Ar}{V} & 10 & 45 \\
\ion{O}{I} & 1 & 0 & \ion{Ar}{VI} & 9 & 36 \\
\ion{O}{II} & 36 & 630 & \ion{Ar}{VII} & 1 & 0 \\
\ion{O}{III} & 33 & 528 & \ion{K}{III} & 10 & 45 \\
\ion{O}{IV} & 29 & 406 & \ion{K}{IV} & 23 & 253 \\
\ion{O}{V} & 54 & 1431 & \ion{K}{V} & 19 & 171 \\
\ion{O}{VI} & 35 & 595 & \ion{K}{VI} & 1 & 0 \\
\ion{O}{VII} & 1 & 0 & \ion{Ca}{II} & 1 & 0 \\
\ion{Ne}{II} & 10 & 45 & \ion{Ca}{III} & 14 & 91 \\
\ion{Ne}{III} & 18 & 153 & \ion{Ca}{IV} & 24 & 276 \\
\ion{Ne}{IV} & 35 & 595 & \ion{Ca}{V} & 15 & 105 \\
\ion{Ne}{V} & 54 & 1431 & \ion{Ca}{VI} & 15 & 105 \\
\ion{Ne}{VI} & 49 & 1176 & \ion{Ca}{VII} & 20 & 190 \\
\ion{Ne}{VII} & 1 & 0 & \ion{Ca}{VIII} & 1 & 0 \\
\ion{Na}{III} & 26 & 325 & \ion{Fe}{I} & 13 & 40 \\
\ion{Na}{IV} & 14 & 91 & \ion{Fe}{II} & 14 & 48 \\
\ion{Na}{V} & 1 & 0 & \ion{Fe}{III} & 13 & 40 \\
\ion{Mg}{II} & 1 & 0 & \ion{Fe}{IV} & 18 & 77 \\
\ion{Mg}{III} & 43 & 903 & \ion{Fe}{V} & 22 & 107 \\
\ion{Mg}{IV} & 1 & 0 & \ion{Fe}{VI} & 29 & 194 \\
\ion{Al}{III} & 1 & 0 & \ion{Fe}{VII} & 19 & 87 \\
\ion{Al}{IV} & 10 & 45 & \ion{Fe}{VIII} & 15 & 54 \\
\ion{Al}{V} & 1 & 0 & \ion{Fe}{IX} & 16 & 62 \\
\ion{Si}{III} & 24 & 276 & \ion{Fe}{X} & 1 & 0 \\
\ion{Si}{IV} & 55 & 1485 &  &  &  \\
\hline
\end{tabular}
\label{tab: line_list}
\end{table}

\onecolumn
\begin{landscape}
\centering
\small  
\begin{longtable}{c c c c c c c c c c c c c c}
\caption{Predicted wind properties from  $\texttt{PoWR}^\textsc{hd}$ models for input stellar mass and temperature.}
\label{tab: wind_properties} \\
\hline
$M_\star/M_\odot$ & $T_\star^\dag$ & $T_{\rm eff}(\tau_\mathrm{R}=2/3)^\dag$ & $\log(\dot{M})^\blacktriangle$ & $\varv_\infty\,[\mathrm{km\,s^{-1}}]$ & $\eta$ & $\tau_{F}(r_\mathrm{sonic})$ & $\Gamma_\mathrm{e}$ & $R_\star/R_{\odot}$ & $R_{\rm crit}/R_{\odot}$ & $R(\tau_\mathrm{R} = 2/3)/R_\odot$ & $\varv_\infty/\varv_{\rm esc}(R_\mathrm{crit})$ & $\log (Q_\ion{H}{I})^\star$ & $\log (Q_\ion{He}{II})^\star$ \\
\hline
\endfirsthead
\multicolumn{14}{c}{\tablename\ \thetable\ -- \textit{Continued from previous page}} \\
\hline
$M_\star/M_\odot$ & $T_\star^\dag$ & $T_{\rm eff}(\tau_\mathrm{R}=2/3)^\dag$ & $\log(\dot{M})^\blacktriangle$ & $\varv_\infty\,[\mathrm{km\,s^{-1}}]$ & $\eta$ & $\tau_{F}(r_\mathrm{sonic})$ & $\Gamma_\mathrm{e}$ & $R_\star/R_{\odot}$ & $R_{\rm crit}/R_{\odot}$ & $R(\tau_\mathrm{R} = 2/3)/R_\odot$ & $\varv_\infty/\varv_{\rm esc}(R_\mathrm{crit})$ & $\log (Q_\ion{H}{I})^\star$ & $\log (Q_\ion{He}{II})^\star$ \\
\hline
\endhead
\hline
\multicolumn{14}{r}{\textit{Continued on next page}} \\
\endfoot
\hline
\endlastfoot
75 & 50 & 48.60 & $-5.921$ & 3950.00 & 0.23 & 0.37 & 0.35 & 13.36 & 16.13 & 14.15 & 2.97 & 49.86 & 45.13 \\
65 & 50 & 48.10 & $-5.767$ & 3167.06 & 0.27 & 0.45 & 0.40 & 13.36 & 16.52 & 14.44 & 2.59 & 49.86 & 44.05 \\
55 & 50 & 46.87 & $-5.544$ & 2474.46 & 0.35 & 0.67 & 0.48 & 13.36 & 17.85 & 15.19 & 2.29 & 49.86 & 41.38 \\
50 & 50 & 45.70 & $-5.407$ & 2069.60 & 0.40 & 0.85 & 0.52 & 13.36 & 18.84 & 15.99 & 2.06 & 49.86 & 41.56 \\
45 & 50 & 41.91 & $-5.066$ & 1476.36 & 0.62 & 1.63 & 0.58 & 13.36 & 20.53 & 18.90 & 1.62 & 49.85 & 41.16 \\
40 & 50 & 32.43 & $-4.573$ & 960.35 & 1.26 & 4.35 & 0.66 & 13.36 & 22.61 & 31.64 & 1.17 & 49.79 & 38.79 \\
37 & 50 & 23.99 & $-4.185$ & 811.41 & 2.61 & 12.12 & 0.71 & 13.36 & 22.54 & 57.70 & 1.03 & 49.58 & 34.05 \\
95 & 45 & 44.03 & $-5.971$ & 3986.11 & 0.21 & 0.37 & 0.28 & 16.50 & 19.88 & 17.23 & 2.96 & 49.83 & 44.22 \\
85 & 45 & 43.80 & $-5.802$ & 3222.88 & 0.25 & 0.44 & 0.31 & 16.50 & 19.78 & 17.41 & 2.52 & 49.83 & 41.52 \\
75 & 45 & 43.46 & $-5.723$ & 2954.86 & 0.28 & 0.49 & 0.35 & 16.50 & 20.39 & 17.69 & 2.50 & 49.83 & 41.51 \\
65 & 45 & 42.86 & $-5.629$ & 2731.49 & 0.32 & 0.53 & 0.40 & 16.50 & 21.25 & 18.18 & 2.53 & 49.83 & 41.31 \\
60 & 45 & 42.24 & $-5.483$ & 2156.62 & 0.35 & 0.72 & 0.44 & 16.50 & 22.39 & 18.70 & 2.14 & 49.84 & 41.25 \\
55 & 45 & 41.27 & $-5.363$ & 1915.75 & 0.41 & 0.84 & 0.48 & 16.50 & 23.13 & 19.57 & 2.02 & 49.83 & 40.93 \\
50 & 45 & 38.92 & $-5.079$ & 1279.83 & 0.52 & 1.43 & 0.53 & 16.50 & 24.33 & 22.01 & 1.45 & 49.82 & 40.15 \\
45 & 45 & 32.56 & $-4.643$ & 950.89 & 1.06 & 3.69 & 0.58 & 16.50 & 25.56 & 31.43 & 1.16 & 49.77 & 38.45 \\
40 & 45 & 23.50 & $-4.271$ & 766.46 & 2.02 & 7.99 & 0.66 & 16.50 & 26.83 & 60.28 & 1.02 & 49.57 & 34.89 \\
38 & 45 & 18.70 & $-3.996$ & 721.95 & 3.58 & 19.62 & 0.69 & 16.50 & 25.44 & 95.31 & 0.96 & 49.08 & 30.51 \\
95 & 40 & 38.84 & $-5.935$ & 3576.52 & 0.20 & 0.34 & 0.28 & 20.88 & 26.60 & 22.14 & 3.07 & 49.79 & 40.82 \\
85 & 40 & 38.52 & $-5.712$ & 2575.98 & 0.25 & 0.51 & 0.31 & 20.88 & 27.06 & 22.50 & 2.36 & 49.80 & 40.26 \\
75 & 40 & 38.04 & $-5.589$ & 2082.12 & 0.26 & 0.60 & 0.35 & 20.88 & 28.44 & 23.09 & 2.08 & 49.80 & 40.00 \\
70 & 40 & 37.54 & $-5.513$ & 1699.67 & 0.26 & 0.66 & 0.37 & 20.88 & 29.34 & 23.68 & 1.79 & 49.79 & 39.74 \\
65 & 40 & 37.17 & $-5.477$ & 1702.40 & 0.28 & 0.64 & 0.40 & 20.88 & 29.67 & 24.14 & 1.87 & 49.79 & 39.46 \\
60 & 40 & 36.17 & $-5.306$ & 1326.53 & 0.32 & 0.87 & 0.44 & 20.88 & 31.28 & 25.47 & 1.55 & 49.78 & 39.11 \\
55 & 40 & 33.97 & $-5.057$ & 1089.28 & 0.47 & 1.53 & 0.48 & 20.88 & 32.57 & 28.89 & 1.36 & 49.77 & 38.69 \\
50 & 40 & 28.50 & $-4.643$ & 922.14 & 1.03 & 3.75 & 0.52 & 20.88 & 32.47 & 40.89 & 1.21 & 49.68 & 37.28 \\
45 & 40 & 21.71 & $-4.214$ & 674.13 & 2.03 & 8.40 & 0.58 & 20.88 & 30.26 & 70.63 & 0.90 & 49.43 & 31.57 \\
40 & 40 & 15.26 & $-3.882$ & 647.85 & 4.18 & 23.07 & 0.65 & 20.88 & 31.80 & 142.09 & 0.94 & 48.20 & 29.62 \\
115 & 35 & 33.97 & $-6.220$ & 4584.31 & 0.14 & 0.23 & 0.22 & 27.27 & 37.80 & 28.95 & 4.26 & 49.66 & 39.12 \\
105 & 35 & 33.80 & $-5.826$ & 2407.59 & 0.18 & 0.36 & 0.25 & 27.27 & 35.29 & 29.23 & 2.26 & 49.68 & 38.84 \\
95 & 35 & 33.55 & $-5.749$ & 2150.50 & 0.19 & 0.38 & 0.27 & 27.27 & 35.92 & 29.66 & 2.15 & 49.69 & 38.70 \\
85 & 35 & 33.14 & $-5.623$ & 1743.49 & 0.20 & 0.43 & 0.30 & 27.27 & 37.14 & 30.39 & 1.87 & 49.69 & 38.52 \\
75 & 35 & 32.38 & $-5.479$ & 1483.89 & 0.24 & 0.52 & 0.34 & 27.27 & 39.54 & 31.82 & 1.75 & 49.68 & 38.36 \\
65 & 35 & 30.88 & $-5.280$ & 1300.24 & 0.34 & 0.77 & 0.40 & 27.27 & 43.80 & 34.93 & 1.73 & 49.67 & 38.21 \\
60 & 35 & 29.35 & $-5.021$ & 1107.27 & 0.52 & 1.46 & 0.43 & 27.27 & 46.64 & 38.73 & 1.58 & 49.64 & 37.68 \\
55 & 35 & 26.78 & $-4.786$ & 850.25 & 0.68 & 2.38 & 0.47 & 27.27 & 49.88 & 46.22 & 1.31 & 49.57 & 36.44 \\
50 & 35 & 20.39 & $-4.391$ & 610.33 & 1.22 & 5.25 & 0.53 & 27.27 & 46.74 & 79.55 & 0.96 & 49.34 & 33.20 \\
47 & 35 & 16.98 & $-4.092$ & 627.82 & 2.50 & 10.65 & 0.55 & 27.27 & 41.94 & 115.41 & 0.96 & 48.69 & 30.61 \\
45 & 35 & 14.39 & $-3.937$ & 588.23 & 3.35 & 14.63 & 0.58 & 27.27 & 41.10 & 158.68 & 0.91 & 47.70 & 28.50 \\
105 & 30 & 28.45 & $-6.071$ & 2752.83 & 0.11 & 0.21 & 0.23 & 37.12 & 57.54 & 41.27 & 3.31 & 49.45 & 37.15 \\
95 & 30 & 28.06 & $-5.627$ & 1016.91 & 0.12 & 0.38 & 0.26 & 37.12 & 56.39 & 42.40 & 1.27 & 49.50 & 36.45 \\
85 & 30 & 27.41 & $-5.463$ & 818.44 & 0.14 & 0.49 & 0.29 & 37.12 & 59.62 & 44.37 & 1.11 & 49.49 & 36.05 \\
75 & 30 & 26.20 & $-5.330$ & 652.43 & 0.15 & 0.70 & 0.33 & 37.12 & 71.68 & 48.60 & 1.03 & 49.46 & 35.17 \\
70 & 30 & 25.76 & $-5.178$ & 628.78 & 0.21 & 0.75 & 0.35 & 37.12 & 66.11 & 50.34 & 0.99 & 49.41 & 35.19 \\
65 & 30 & 23.70 & $-5.021$ & 555.10 & 0.26 & 1.17 & 0.38 & 37.12 & 82.56 & 59.26 & 1.01 & 49.38 & 34.16 \\
60 & 30 & 20.40 & $-4.788$ & 652.21 & 0.52 & 2.47 & 0.41 & 37.12 & 112.18 & 79.78 & 1.45 & 49.20 & 32.72 \\
55 & 30 & 19.30 & $-4.559$ & 619.04 & 0.84 & 3.01 & 0.45 & 37.12 & 91.87 & 89.71 & 1.30 & 49.08 & 32.80 \\
50 & 30 & 13.96 & $-4.153$ & 534.49 & 1.85 & 12.54 & 0.49 & 37.12 & 142.91 & 170.41 & 1.47 & 47.63 & 28.56 \\
47 & 30 & 10.70 & $-3.712$ & 467.63 & 4.46 & 21.16 & 0.56 & 37.12 & 54.16 & 291.46 & 0.81 & 39.01 & 21.32 \\
105 & 25 & 23.06 & $-6.866$ & 7239.04 & 0.05 & 0.07 & 0.21 & 53.45 & 114.60 & 62.79 & 12.27 & 48.96 & 36.41 \\
95 & 25 & 22.81 & $-6.043$ & 987.47 & 0.04 & 0.14 & 0.25 & 53.45 & 110.65 & 64.21 & 1.73 & 49.03 & 34.29 \\
85 & 25 & 22.06 & $-5.664$ & 665.43 & 0.07 & 0.21 & 0.29 & 53.45 & 107.82 & 68.62 & 1.22 & 49.02 & 33.55 \\
80 & 25 & 21.72 & $-5.425$ & 662.36 & 0.12 & 0.34 & 0.30 & 53.45 & 101.29 & 70.72 & 1.21 & 48.94 & 33.30 \\
75 & 25 & 21.02 & $-5.253$ & 680.42 & 0.19 & 0.47 & 0.32 & 53.45 & 108.08 & 75.40 & 1.33 & 48.92 & 33.05 \\
65 & 25 & 18.38 & $-4.791$ & 684.29 & 0.54 & 1.48 & 0.37 & 53.45 & 138.98 & 98.44 & 1.62 & 48.63 & 30.75 \\
60 & 25 & 16.27 & $-4.591$ & 655.52 & 0.83 & 2.11 & 0.41 & 53.45 & 164.64 & 125.36 & 1.76 & 48.44 & 30.38 \\
55 & 25 & 14.67 & $-4.269$ & 515.87 & 1.37 & 3.40 & 0.44 & 53.45 & 157.05 & 155.06 & 1.41 & 46.86 & 26.90 \\
53 & 25 & 12.62 & $-3.973$ & 440.40 & 2.31 & 10.08 & 0.46 & 53.45 & 159.13 & 209.30 & 1.24 & 39.32 & 18.49 \\
105 & 23 & 21.15 & $-5.895$ & 734.14 & 0.05 & 0.16 & 0.23 & 63.15 & 117.47 & 74.67 & 1.26 & 48.68 & 33.02 \\
95 & 23 & 20.77 & $-5.595$ & 671.28 & 0.08 & 0.24 & 0.25 & 63.15 & 109.57 & 77.39 & 1.17 & 48.67 & 32.54 \\
85 & 23 & 20.00 & $-5.416$ & 684.97 & 0.13 & 0.30 & 0.28 & 63.15 & 120.06 & 83.26 & 1.32 & 48.63 & 32.21 \\
75 & 23 & 19.19 & $-5.144$ & 712.98 & 0.25 & 0.51 & 0.32 & 63.15 & 125.17 & 90.38 & 1.49 & 48.66 & 32.02 \\
65 & 23 & 16.59 & $-4.630$ & 639.26 & 0.74 & 1.64 & 0.37 & 63.15 & 157.51 & 120.45 & 1.61 & 48.20 & 29.19 \\
60 & 23 & 14.84 & $-4.371$ & 517.38 & 1.08 & 2.68 & 0.41 & 63.15 & 170.26 & 150.96 & 1.41 & 47.16 & 26.11 \\
55 & 23 & 12.01 & $-3.954$ & 423.68 & 2.32 & 6.08 & 0.44 & 63.15 & 164.77 & 231.06 & 1.19 & 39.16 & 19.71 \\
115 & 20 & 18.36 & $-5.528$ & 645.62 & 0.09 & 0.26 & 0.21 & 83.52 & 137.64 & 98.96 & 1.15 & 48.21 & 30.92 \\
105 & 20 & 17.99 & $-5.258$ & 662.75 & 0.18 & 0.43 & 0.23 & 83.52 & 140.23 & 103.27 & 1.24 & 48.20 & 30.21 \\
95 & 20 & 17.24 & $-4.968$ & 636.02 & 0.34 & 0.87 & 0.25 & 83.52 & 156.27 & 112.21 & 1.32 & 48.10 & 29.09 \\
85 & 20 & 16.67 & $-4.806$ & 561.64 & 0.43 & 1.12 & 0.28 & 83.52 & 164.54 & 119.68 & 1.27 & 47.83 & 27.55 \\
75 & 20 & 15.29 & $-4.475$ & 416.77 & 0.69 & 1.88 & 0.32 & 83.52 & 171.80 & 142.48 & 1.02 & 43.73 & 26.17 \\
65 & 20 & 12.04 & $-4.083$ & 396.90 & 1.61 & 3.93 & 0.37 & 83.52 & 186.67 & 230.14 & 1.09 & 39.12 & 20.93 \\
60 & 20 & 9.20 & $-3.827$ & 393.05 & 2.88 & 7.22 & 0.40 & 83.52 & 191.35 & 392.57 & 1.14 & 38.62 & 21.08 \\
57 & 20 & 7.65 & $-3.641$ & 390.15 & 4.39 & 17.42 & 0.43 & 83.52 & 193.77 & 564.04 & 1.17 & 38.40 & 20.50 \\
115 & 17 & 14.93 & $-6.013$ & 446.27 & 0.02 & 0.07 & 0.21 & 115.60 & 366.11 & 149.57 & 1.29 & 47.25 & 26.42 \\
105 & 17 & 14.61 & $-5.816$ & 395.52 & 0.03 & 0.10 & 0.23 & 115.60 & 436.04 & 156.13 & 1.31 & 47.09 & 25.79 \\
95 & 17 & 13.85 & $-5.536$ & 360.42 & 0.05 & 0.15 & 0.25 & 115.60 & 554.76 & 173.28 & 1.41 & 46.70 & 26.20 \\
85 & 17 & 12.65 & $-5.046$ & 357.23 & 0.16 & 0.36 & 0.27 & 115.60 & 503.67 & 207.60 & 1.41 & 41.43 & 22.17 \\
75 & 17 & 11.43 & $-4.557$ & 391.62 & 0.53 & 1.25 & 0.31 & 115.60 & 499.39 & 254.68 & 1.64 & 39.79 & 21.36 \\
65 & 17 & 7.89 & $-3.801$ & 389.23 & 3.03 & 9.43 & 0.35 & 115.60 & 482.28 & 534.23 & 1.72 & 38.40 & 20.56 \\
115 & 15 & 12.78 & $-6.468$ & 376.41 & 0.01 & 0.01 & 0.20 & 148.48 & 964.83 & 204.10 & 1.77 & 46.00 & 26.18 \\
110 & 15 & 12.75 & $-6.176$ & 350.56 & 0.01 & 0.02 & 0.21 & 148.48 & 984.14 & 205.51 & 1.70 & 45.87 & 25.78 \\
105 & 15 & 12.53 & $-5.727$ & 340.79 & 0.03 & 0.08 & 0.22 & 148.48 & 862.53 & 212.71 & 1.58 & 45.46 & 25.90 \\
100 & 15 & 12.31 & $-5.417$ & 351.03 & 0.07 & 0.14 & 0.23 & 148.48 & 658.96 & 220.37 & 1.46 & 42.96 & 25.33 \\
95 & 15 & 12.02 & $-5.095$ & 370.56 & 0.15 & 0.28 & 0.24 & 148.48 & 526.07 & 231.12 & 1.41 & 40.56 & 22.28 \\
85 & 15 & 10.73 & $-4.583$ & 459.92 & 0.59 & 1.01 & 0.27 & 148.48 & 566.01 & 289.79 & 1.93 & 43.77 & 24.13 \\
75 & 15 & 8.93 & $-4.065$ & 404.60 & 1.71 & 2.87 & 0.30 & 148.48 & 562.89 & 415.34 & 1.80 & 38.68 & 18.52 \\
70 & 15 & 7.17 & $-3.713$ & 396.97 & 3.78 & 8.25 & 0.32 & 148.48 & 539.58 & 647.00 & 1.79 & 38.14 & 18.52 \\
135 & 12 & 9.99 & $-4.961$ & 404.03 & 0.22 & 0.38 & 0.16 & 232.00 & 674.20 & 333.80 & 1.46 & 39.40 & 20.62 \\
125 & 12 & 9.65 & $-4.727$ & 522.79 & 0.48 & 0.70 & 0.18 & 232.00 & 657.73 & 357.70 & 1.95 & 40.75 & 21.60 \\
115 & 12 & 9.26 & $-4.568$ & 414.61 & 0.55 & 0.86 & 0.19 & 232.00 & 725.24 & 387.29 & 1.69 & 38.87 & 20.92 \\
105 & 12 & 8.38 & $-4.222$ & 407.55 & 1.20 & 1.82 & 0.21 & 232.00 & 779.76 & 474.46 & 1.80 & 38.29 & 19.46 \\
95 & 12 & 7.24 & $-3.916$ & 424.64 & 2.53 & 3.66 & 0.23 & 232.00 & 793.68 & 634.59 & 1.99 & 39.24 & 19.53 \\
85 & 12 & 6.21 & $-3.627$ & 374.62 & 4.35 & 5.95 & 0.26 & 232.00 & 752.85 & 865.56 & 1.81 & 38.02 & 19.04 \\
\end{longtable}

\tablefoot{$^\dag$ Temperatures $T_\star$ and $T_{\rm eff}(\tau_\mathrm{R}=2/3)$ are in kK.\\
$^\blacktriangle$ Mass-loss rate $\dot{M}$ is in $M_\odot\, \mathrm{yr^{-1}}$.\\ 
$^\star$ $Q_\ion{H}{I}$ and $Q_\ion{He}{II}$ are expressed in number of photons per second.}
\end{landscape}

\end{document}